\RequirePackage{xcolor}
\documentclass[9pt,shortpaper,twoside,web]{ieeecolor}
\usepackage{generic}
\usepackage{latexsym}
\usepackage{graphicx}
\usepackage{amsfonts,amssymb,amsmath}
\usepackage{cite}
\usepackage{algorithmic}
\usepackage{graphicx}
\usepackage{textcomp}
\usepackage{booktabs}
\usepackage{pgfplots}
\usepackage{multirow}
\usepackage{float}
\usepackage{mathrsfs}
\usepackage[ruled]{algorithm2e}
\usepackage{etoolbox}

\usepackage{soul}
\usepackage{booktabs}
\usepackage{xcolor}
\usepackage{lipsum}
\usepackage{xr}
\usepackage[switch]{lineno}

\makeatletter
\newcommand*{\addFileDependency}[1]{
  \typeout{(#1)}
  \@addtofilelist{#1}
  \IfFileExists{#1}{}{\typeout{No file #1.}}
}
\makeatother

\newcommand*{\myexternaldocument}[1]{
    \externaldocument{#1}
    \addFileDependency{#1.tex}
    \addFileDependency{#1.aux}
}

\myexternaldocument{supplementary}
\SetCommentSty{mycommfont}
\newcommand{\etal}{\textit{et al.}}
\newcommand{\ie}{i.e.}

{\color{yellow}}%
{}
\newcommand{\hlo}[1]{#1}
\newcommand{\fhlo}[1]{#1}
\newcommand{\coloro}[1]{#1}

\newcommand{\url}[1]{{#1}}

\definecolor{newcolor}{rgb}{.8,.349,.1}
\sethlcolor{yellow}
\usepackage{arydshln}

\def\BibTeX{{\rm B\kern-.05em{\sc i\kern-.025em b}\kern-.08em T\kern-.1667em\lower.7ex\hbox{E}\kern-.125emX}}
\begin{document}
	\title{Cross-Modality Multi-Atlas Segmentation \fhlo{via Deep Registration and Label Fusion}}
	\author{Wangbin Ding, Lei Li, Xiahai Zhuang*, and Liqin Huang*
		\thanks{Corresponding authors: Xiahai Zhuang; Liqin Huang. 
			Xiahai Zhuang and Liqin Huang are co-senior authors and contribute equally. 
			\fhlo{This work was supported by the National Nature Science Foundation of China (62011540404, 62111530195, and 61971142), Fujian Provincial Natural Science Foundation Project (2021J02019, 2021J01578 and 2019Y9070), and Fuzhou Science and Technology Project (2020-GX-17).}
		}
		\thanks{Liqin Huang and Wangbin Ding are with the College of Physics and Information Engineering, Fuzhou University, Fuzhou 350117, China (e-mail: n191110003@fzu.edu.cn; hlq@fzu.edu.cn).}
		\thanks{Xiahai Zhuang is with the School of Data Science, Fudan University, Shanghai 200433, China (e-mail: zxh@fudan.edu.cn).}
		\thanks{Lei Li is with the School of Biomedical Engineering, Shanghai Jiao Tong University, Shanghai 200230, China (e-mail: lilei.sky@sjtu.edu.cn).}
	}
	
	\maketitle
	
	\begin{abstract}
		Multi-atlas segmentation (MAS) is a promising framework for medical image segmentation. Generally, MAS methods register multiple atlases, i.e.,  medical images with corresponding labels, to a target image; and the transformed atlas labels can be combined to generate target segmentation via label fusion schemes. Many conventional MAS methods employed the atlases from the same modality as the target image.	However, the number of atlases with the same modality may be limited or even missing in many clinical applications. Besides, conventional MAS methods suffer from the computational burden of registration or label fusion procedures. In this work, we design a novel cross-modality MAS framework, which uses available atlases from a certain modality to segment a target image from another modality. To boost the computational efficiency of the framework, both the image registration and label fusion are achieved by well-designed deep neural networks. For the atlas-to-target image registration, we propose a bi-directional registration network (BiRegNet), which can efficiently align images from different modalities. For the label fusion, we design a similarity estimation network (SimNet), which estimates the fusion weight of each atlas by measuring its similarity to the target image. SimNet can learn multi-scale information for similarity estimation to improve the performance of label fusion. The proposed framework was evaluated by the left ventricle and liver segmentation tasks on the MM-WHS and CHAOS datasets, respectively. Results have shown that the framework is effective for cross-modality MAS in both registration and label fusion\fhlo{\footnote{\fhlo{The code has been released at https://github.com/NanYoMy/cmmas}}}.
	\end{abstract}

	\section{Introduction}

	\IEEEPARstart{S}{egmentation} is an essential step in medical image analysis. Many clinical applications rely on segmentation to extract specific anatomy or to compute certain functional indices. For instance, anatomy structural segmentation is a general prerequisite in clinical cardiology \cite{jour/jbhi/xue2021left} and hepatology \cite{jour/jbhi/ji2013acm} applications.

	Multi-atlas segmentation (MAS) is one of most successful  techniques for the medical image segmentation \cite{journal/mia/iglesias2015multi}. 
	Generally,  MAS methods mainly contain two steps, i.e., 
	the atlas-to-target registration and the label fusion of multiple warped atlases.
	The registration aims to warp the atlases to the target image space. Consequently, the labels of the warped atlases are regarded as the candidate segmentations of the target image. This step is therefore considered as an effective scheme in incorporating the prior shape knowledge from the atlases \cite{journal/mia/antonelli2019gas}. 
	The label fusion is designed to jointly fuse the multiple candidates into one unified result for the final segmentation of the target image. This joint decision making of MAS considering multiple candidates can improve the generalization capability and robustness, comparing with single-atlas segmentation \cite{journal/PAMI/kittler1998combining}.

\begin{figure}[htp]
		\includegraphics[width=\columnwidth]{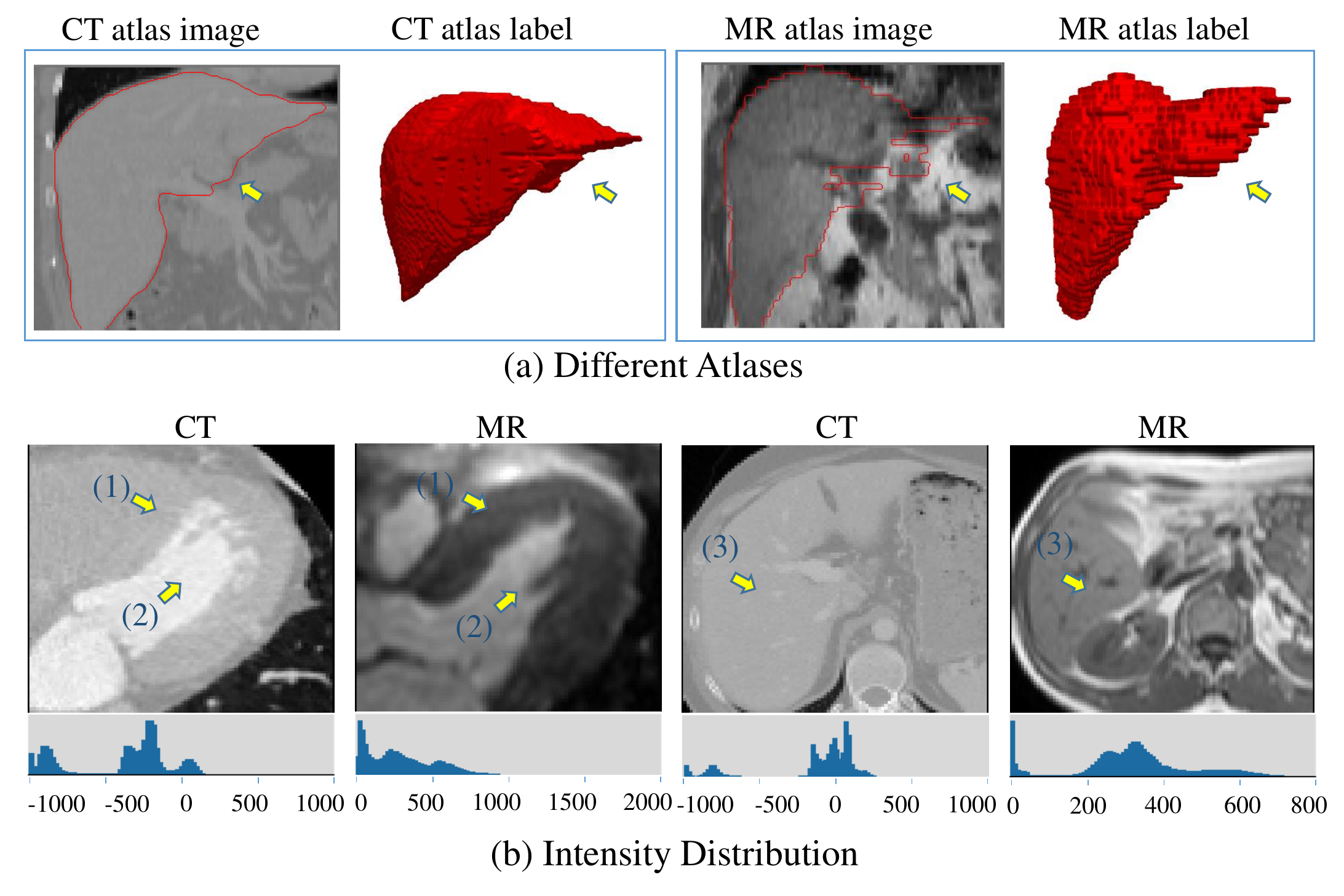}
		\caption{(a) Examples of computed tomography (CT) and magnetic resonance (MR) atlas. As indicated by purple arrows, the label of MR atlas provides relatively poor structural information due to its low coronal resolution. 
		(b) Examples of CT and MR image with corresponding intensity distribution. Arrow (1), (2) and (3) point to left ventricle myocardium (Myo), left ventricle blood cavity (LVC) and liver, respectively. The appearance of left ventricle (LV) and liver are largely different in CT and MR images.}
		\label{fig:0}    
	\end{figure}

\hlo{Most of the current MAS methods use single-modality atlases, whose intensity images have the same modality as the target images \cite{journal/tmi/Tang2019}. In clinical practice, multiple modalities of medial images are usually acquired to assess different properties of the same region of interest \cite{joural/mia/kavur2020chaos}. There are many benefits to analyze target images by using information from another modality (cross-modality). 
For instance, Li \etal \cite{jour/mia/li2020atrial} employed the structural information from anatomical MRIs to assist the segmentation of atrial scars on LGE MRIs. By developing a cross-modality MAS method, one can also employ high-quality images (such as CT, see left of Fig. \ref{fig:0} (a)) to improve the segmentation of low-quality ones (such as MR, see right of Fig. \ref{fig:0} (a)). Meanwhile, cross-modality MAS methods could bring convenience when performing multi-modality segmentation \cite{journal/MIA/EugenioIglesias2013}. For example, multi-modality abdominal images, \ie, T1-DUAL, T2-SPIR and CT, are usually acquired for liver analysis \cite{joural/mia/kavur2020chaos}. Based on a cross-modality MAS method, we can perform segmentation of multi-modality images (such as T1-DUAL and T2-SPIR) by only using atlases from a single high-quality modality (such as CT). Nevertheless, the intensity distributions of an anatomy can vary largely among different modalities (see Fig. \ref{fig:0} (b)), which poses additional challenges for the cross-modality MAS. Recently,} several works have been proposed to adopt cross-modality atlases in MAS framework \cite{conf/icip/kasiri2014cross,journal/MIA/EugenioIglesias2013}, which however are generally computationally expensive. The reason has two folds. First, these methods perform the registration step in an iterative fashion. Second, the label fusion is generally processed patch-wisely via local information \cite{jour/tmi/bai2013probabilistic,journal/mia/sanroma2018lf,conf/miccai/coup2010}. In summary, further effort is required to explore the application of cross-modality atlases as well as to develop a computationally efficient MAS framework.	
	
	\begin{figure*}[ht]
		\centering
		\includegraphics[width=1.0\textwidth]{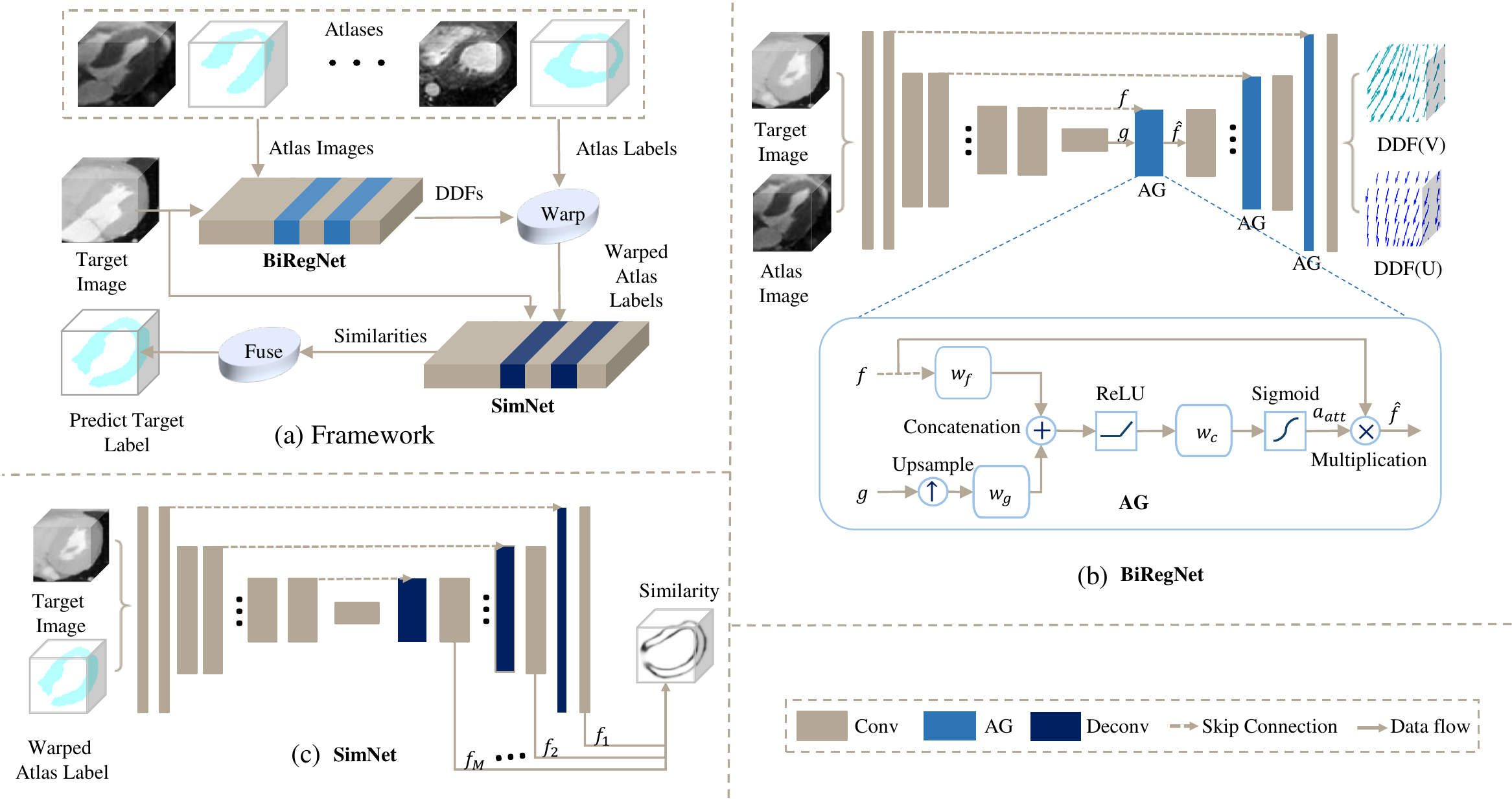}
		\caption{The workflow and network architectures of the proposed method. 
			(a) the pipeline of the proposed cross-modality MAS framework. 
			BiRegNet first predicts DDFs between the atlas images and the target image. 
			Then, the atlas labels can be warped by the predicted DDFs as the candidate segmentations of the target image. 
			Next, each warped atlas label (candidate segmentation) is weighted according to its similarity to the target image via SimNet. 
			Finally, the warped atlas labels are fused according to similarity weights for the final segmentation of target image.
			\hlo{(b) the architecture of BiRegNet and AG module. For the details of $g$, $f$, $w_g$, $w_f$, $w_c$, $a_{att}$ and $\hat{f}$, please refer to \eqref{eq:att} and \eqref{equ:ag_out}.}
			(c) the architecture of SimNet.
			Note that these figures take Myo segmentation as an example, while liver and LV are processed in the same way. \hlo{Detail architectures of BiRegNet and SimNet can be found in the Supplementary Material.}
		}
		\label{fig:framework}
	\end{figure*}

	In this work, we propose a novel deep learning-based cross-modality MAS framework, which aims to segment a target image from modality $\mathcal{A}$ by using a set of atlases from modality $\mathcal{B}$ (cross-modality atlases). 
	To achieve atlas-to-target registration, we propose a bi-directional cross-modality registration network (BiRegNet), which can jointly estimate the forward and backward dense displacement fields (DDFs). 
	We further constrain the forward and backward DDFs by introducing a consistency loss.
	Additionally, attention gate (AG) \cite{arxiv/Oktay} module is embedded into BiRegNet to emphasize misaligned areas for better registration performance.  
	For label fusion, we introduce a similarity estimation network (SimNet) to estimate similarity-based fusion weight for each warped atlas. Specifically, SimNet is particularly designed to compute similarities between target image and warped atlas labels instead of warped atlas images. This is because using warped atlas labels could improve the accuracy of the similarity estimation \cite{journal/mp/Zhuang2015}.
	In this way, both the registration and label fusion steps are achieved by deep neural networks (DNNs), i.e., BiRegNet and SimNet. 
	Notably, the proposed MAS framework has better computational efficiency compared to conventional MAS, thanks to the well-designed DNN schemes.
	Moreover, the accuracy of the proposed MAS framework is validated by two public datasets, \ie, the MM-WHS dataset of MICCAI 2017 \cite{journal/mia/zhuang2019evaluation} and the  CHAOS dataset of ISBI 2019 \cite{joural/mia/kavur2020chaos}. 
	This paper extends a preliminary version of work presented at  MICCAI 2020 \cite{conf/miccai/ding2020cross}. 
	First, we expand the architecture of the similarity estimation networks. Second, we employ more data to verify the effectiveness of the proposed framework. Finally, more objective analyses and studies are added.

	The remainder of this paper is organized as follows: 
	Section \ref{sec:reviws} reviews the related work. 
	Section \ref{sec:reg} and \ref{sec:sim} describe the details of the proposed cross-modality MAS framework. 
	Section \ref{sec:exp} presents the experiments and results, followed by the discussion and conclusion in Section \ref{sec:diss}.

	\section{Related Work}
	\label{sec:reviws}
	
	\subsection{MAS and Label Fusion}

	The principle of MAS framework is propagating atlas labels to the target image coordinate, and the propagated labels can be fused to generate the target image label. 
	Here, the fusion of multiple atlas labels is the key component of MAS framework. 
	Majority voting (MV) is one of the most straightforward fusion strategies. 
	It assumes that each atlas contributes equally. 
	In contrast, local weighted voting methods can assign adaptive weights to atlases based on their similarity to the target image. 
	Coup{\'e} \etal \cite{conf/miccai/coup2010} proposed a patch-based weighted label fusion method, where the weight of each atlas was calculated by the intensity similarity between local patches.  
	Zhuang \& Shen \cite{journal/mia/zhuang2016mmm} proposed a metric of conditional probability of images to measure the pixel-wise similarities between multi-modality atlases and target images, and applied it to the local weighted label fusion of whole heart CT-MR images in the MAS.  
	Besides, based on the probability theory, Warfield \etal \cite{journal/tmi/warfield2004simultaneous} developed a expectation-maximization algorithm for label fusion. 
	The algorithm could simultaneously compute a probabilistic estimation of target image labels and a measure of the performance level represented by each warped atlas label.
	Wang \etal \cite{jour/pami/wang2012multi} proposed a joint label fusion method, where label fusion was formulated by minimizing the total expectation of segmentation errors.

	Recently, DNNs are employed to further enhance label fusion from three different aspects. 
	The first is to select reliable atlases for fusion. 
	For example, Ding \etal \cite {conf/miccai/ding2019votenet} proposed a DNN model to locally select reliable warped atlas labels.  The model enhanced the label fusion performance by suppressing low-quality warped atlas labels. 
	The second is to improve the estimation of fusion weights.
	Sanroma \etal \cite {journal/mia/sanroma2018lf} extracted better DNN-based features from atlas and target image patches for fusion weight calculation. 
	Similarly, Xie \etal \cite {conf/miccai/xie2019msg} developed a DNN to predict the probabilities of an atlas patch and a target patch with the same label. 
	In this way, they achieved a spatial-aware weighted voting for fusion. 
	The last is to upgrade fusion strategies directly. 
	Yang \etal \cite {journal/mia/yang2018neural} parameterized the fusion strategy via a DNN in an end-to-end style. 
	However, the aforementioned DNN-based fusion works mainly focused on mono-modality, and maybe can not be directly applied to cross-modality MAS. 
	Therefore, in this work we explore a cross-modality multi-atlas fusion method based on DNN. 
	
	\subsection{Image Registration}
	
	Registration aims to estimate spatial transformation between image pairs. 
	Typically, the transformation can be used for atlas label propagation in MAS.
	To achieve the registration of cross-modality atlas image to target image, one way is to use modality-invariant metrics as the registration criteria, such as mutual information (MI) \cite{journal/pr/luan2008multimodality}, normalized mutual information (NMI) \cite{journal/pr/studholme1999overlap} and spatially encoded MI \cite{journal/tmi/zhuang2011nonrigid}. 
	However, the MI-based metrics only consider the intensity distribution of images, and ignore the structural information. 
	Therefore, it can not efficiently measure the local anatomical similarity between images \cite{conf/miccai/qin2019unsupervised}.
	An alternative way is to employ structural representations of images. 
	Wachinger \etal \cite{journal/mia/wachinger2012entropy} employed entropy and Laplacian images to represent structure information of original intensity images, and then directly employed ${L_1}$-norm or ${L_2}$-norm criterion for cross-modality image registration. 
	Similarly, Heinrich \etal  \cite{journal/mia/Heinrich2012} designed a modality independent neighborhood descriptor for 3D CT and MR image registration.
	One can see that these approaches usually optimized registrations in an iterative way, which limits their application in the time-sensitive scenarios.

	DNNs have shown great potentials to reduce computational consumption for registration \cite{journal/mva/haskins2020deep}.
	As for the DNN-based registration models, they are mainly based on fully convolutional architectures, and can rapidly compute a transformation for a registration task \cite{journal/mia/Fan2019,journal/ni/Yang2017}. 
	For cross-modality image registrations, Balakrishnan \etal \cite{journal/tmi/Balakrishnan2019} presented a diffeomorphic registration network. 
	It could generate stationary velocity fields (SVFs) for a pair of images. 
	Hu \etal \cite{journal/mia/hu2018weakly} designed a weakly-supervised registration network by employing anatomical labels as supervisions. 
	It is worth noting that the registration networks normally require a regularization term, such as bending energy on DDFs or $L_2$-norm on gradient of SVFs, to ensure the smoothness of the predicated deformation fields. 
	These regularization techniques can globally control the spatial regularity of transformation fields (DDFs or SVFs)  but with limited local flexibility. 
	In contrast, we introduce a new form of consistency constraint, which regularizes the forward and backward DDFs predicted from BiRegNet in a transformation-cycle fashion.
	
	\section{Method}
	\label{sec:method}
	This section introduces the cross-modality MAS framework. 
	As shown in Fig. \ref{fig:framework} (a), the framework mainly includes two steps: Section \ref{sec:reg} presents the registration step, where BiRegNet is fully described; Section \ref{sec:sim} presents the label fusion step, where cross-modality multi-atlas fusion method and SimNet are introduced.

	\subsection{{Registration}}
	\label{sec:reg}
	
	\subsubsection{{BiRegNet}}
	BiRegNet performs registration between each cross-modality atlas image and target image. 
	Let $I_t$ be the target image to be segmented, $L_t$ be the gold standard label of $I_t$, and $\{ (I_a^i,L_a^i )|i=1,...,N \}$ be the cross-modality atlases, where $I_a^i$ and $L_a^i$ are the image and label of $i$-th atlas, respectively.  
	For each pair of $I_a^i$ and  $I_t$, two registrations can be performed by switching the role of $I_a^i$ and $I_t$. 
	We denote the DDF from $I_a^i$ to $I_t$ as $U^i$, and vice versa as $V^i$. 
	For convenience, we abbreviate $I_a^i$, $L_a^i$, $U^i$ and $V^i$ as $I_a$, $L_a$, $U$ and $V$ when no confusion is caused. 
	\hlo{$L_a$, $I_a$, $L_t$ and $I_t$ can be warped by $U$ and $V$ as follows,}
	\begin{equation}
		\left\{ 
		\begin{split}
			&	\tilde{L}_a(x)=L_a (x+U(x)),\\
			&  \tilde{I}_a(x)=I_a (x+U(x)),\\
			&	\tilde{L}_t(x)=L_t (x+V(x)),\\
			&  \tilde{I}_t(x)=I_t (x+V(x)),\\
		\end{split}
		\right. 
	\end{equation}
	\hlo{where $x$ is a spatial point in the  coordinate space $\Omega$; $\tilde{L_a}$, $\tilde{I_a}$, $\tilde{L_t}$ and $\tilde{I_t}$ denote the warped of ${L_a}$, ${I_a}$, ${L_t}$ and ${I_t}$, respectively; $\tilde{L}_a(x)$, $\tilde{I}_a(x)$, $\tilde{L}_t(x)$ and $\tilde{I}_t(x)$ stand for the voxel values of $\tilde{L}_a$, $\tilde{I}_a$, $\tilde{L}_t$ and $\tilde{I}_t$ at point $x$, respectively; and $U(x)$ and $V(x)$ are the displacement vectors at point $x$.}
	Accordingly, BiRegNet is constructed to simultaneously estimate $U$ and $V$ between $I_a$ and $I_t$, which is formulated as follows, 
	\begin{equation}
		U,V=\mathcal{F}_{\theta}(I_a,I_t),
	\end{equation}
	where $\theta$ is the parameter of BiRegNet. 
	The aim of joint estimation is to improve the plausibility of the generated DDFs via a consistency loss (see Section \ref{sec:inv}).

	Fig. \ref{fig:framework} (b) shows the architecture of BiRegNet. 
	The backbone of the network is an U-shape model. 
	It includes four downsample and four upsample modules. 
	On each skip connection of the network, we adapt an AG module to force the network to focus on the informative areas during the registration.
	\hlo{AG is defined as follows, }
	\begin{equation}
		\alpha_{att} =\sigma_{3}(w_{c}\sigma_{2}(w_{g} \sigma_{1}(g)+w_{f}f)),
		\label{eq:att}
	\end{equation}	
	\begin{equation}
		\hat{f}=f\odot\alpha_{att},
		\label{equ:ag_out}
	\end{equation}	
	where $\sigma_{1}$ is the upsample operation using trilinear interpolation, $\sigma_{2}$ is the Relu function, and $\sigma_{3}$ is the Sigmoid function; $w_{f}$, $w_{g}$, and $w_c$ are trainable parameters; 
	$f$ and $g$ denote the feature maps from skip connection and previous convolution layer, respectively; 
	$\alpha_{att}$ is the attention coefficient which can re-calibrate the feature map $f$; 
	$\hat{f}$ is the output of AG module.  
	
	For cross-modality registration, we train BiRegNet by maximizing the similarity, \ie, multi-scale Dice coefficient \cite{journal/mia/hu2018weakly}, between the warped atlas label ($\tilde{L}_a$) and target label ($L_t$). 
	Meanwhile, the network is constructed to jointly predict forward and backward DDFs, thus pairwise registration errors caused by both DDFs should be taken into account. In this way, a symmetric loss function for the network training is designed as follows,
	\begin{equation}
		\mathcal{L}_{Dice}=-\mathcal{D}_{s}(L_t,\tilde{L}_a)-\mathcal{D}_{s}(L_a,\tilde{L}_t),
		\label{eq:dice}
	\end{equation}
	where $\mathcal{D}_{s}(A,B)$ calculates the multi-scale Dice coefficient between label $A$ and $B$.

	\subsubsection{{Consistency Loss}}
	\label{sec:inv}
	The $\mathcal{L}_{Dice}$ only provides a region-level matching criterion to independently estimate a pair of $U$ and $V$. 
	It can not ensure that the pair of DDFs satisfies a correspondence \cite{journal/tmi/christensen2001consis}, \ie,
	the point-to-point mappings defined by $U$ and $V$ are consistent with each other.
	\hlo{Notably, transformation-cycle, which could improve the consistency of a pair of DDFs \cite{journal/tmi/christensen2001consis}, has been introduced in recent registration networks \cite{conf/miccai/kim2019unsupervised,conf/cvpr/mok2020fast,conf/miccai/gu2020pair}. Unlike previous approaches, BiRegNet aims to jointly predict and regularize the forward and backward DDFs, and the transformation consistency is achieved by introducing a constraint loss as follows,}
	\begin{equation}
		\mathcal{L}_{cons}=\sum\limits_{x \in \Omega} \left\|I_{a}^{'}(x)-I_a(x)\right\|_1+\left\|I_{t}^{'}(x)-I_t(x)\right\|_1,
		\label{eq:cons}
	\end{equation}
	where
	$I_{a}^{'}(x)=\tilde{I}_a(x+V(x)),$
	$I_{t}^{'}(x)=\tilde{I}_t(x+U(x)),$
	and $\left\| \cdot  \right\|_1$ calculates the $L_1$-norm. 
	Ideally, $I_{a}^{'}$ (or $I_{t}^{'}$) should be identical to its original image $I_{a}$ (or $I_{t}$) when $U$ and $V$ are consistent to each other.
	Therefore, it can regularize BiRegNet to generate realistic DDFs. 
	
	Finally, by compositing \eqref{eq:dice} and \eqref{eq:cons}, the total trainable loss of BiRegNet is defined as follows,
	\begin{equation}
		\label{equ:reg_loss}
		\mathcal{L}_{reg}=\mathcal{L}_{Dice}+\lambda \mathcal{L}_{cons},
	\end{equation}
	where $\lambda$ is a balance weight between $\mathcal{L}_{Dice}$ and $\mathcal{L}_{cons}$. 
	
	\subsection{{Label Fusion}}
	\label{sec:sim}
	
	\subsubsection{{Fusion Strategy}}
	
	We fuse the warped atlas labels via locally weighted fusion (LWF) strategy \cite {journal/tmi/artaechevarria2009combination}, where each atlas voxel has a different weight in determining the target label.
	Given $\{ (I_a^i,L_a^i )|i=1,...,N \}$, we first register them to $I_t$ by using BiRegNet, and obtain $N$ warped atlases  $\{ (\tilde{I}_a^i,\tilde{L}_a^i )|i=1,...,N \}$.
	Then, LWF can derive the predicted target label $\hat{L}_t$ as follows,
	\begin{equation}
		\label{equ:fusion}
		\hat{L}_t(x)=\mathop{\arg\max}_{l \in \{l_1,l_2,…..l_k\}}  \sum \limits_{i=1}^{N}  W^i (x)\delta(\tilde{L}_a^i(x) \equiv l),
	\end{equation}
	where $x\in \Omega$, $\{l_1,l_2,…..l_k\}$ is a label set, $W^i(x)$ is the fusion weight of the $i$-th atlas at spatial point $x$,  and $\delta(\cdot)$ is the Kronecker delta function. 
	We abbreviate $W^i$ as $W$ when no confusion is caused. 
	In this equation, one can obtain a better $\hat{L}_t$ by providing more appropriate $W$. 
	Thus, we propose a DNN model, \ie, SimNet, to improve the estimation of $W$ for label fusion.

	\subsubsection{{SimNet}}
	SimNet aims to estimate voxel-wise similarities between warped atlases and target image, and the similarities are regarded as the local weights for the fusion of warped atlas labels.
	Normally, conventional LWF methods calculate fusion weights via intensity-based similarity, such as mean square difference of image intensity, within local patches \cite{conf/miccai/coup2010}. 
	However, such similarity is ill-posed when the patches are from different modalities. 
	To address this weakness, Zhuang \etal \cite{journal/mp/Zhuang2015} proposed to calculate \hlo{the conditional entropy between warped atlas labels and target images for multi-modality atlas ranking and selection}, and achieved promising performance.    
	Inspired by this, SimNet is constructed to compute similarity between  $\tilde{L}_a$ and $I_t$ directly. 
	Besides, most existing LWF methods compute similarities patch-wisely \cite{journal/mia/sanroma2018lf,conf/miccai/xie2019msg}, which could achieve promising performance but still time-consuming. 
	To track this problem, SimNet is designed to calculate a voxel-wise similarity map between a pair of $\tilde{L}_a$ and $I_t$ in an one-step manner. \hlo{Note that though SimNet is inspired by Zhuang \etal \cite{journal/mp/Zhuang2015}, we employed a neural network and therefore can estimate similarities in a computationally efficient way.} It is formulated as follows,
	\begin{equation}
		\hat{W}=\mathcal{F}_{\phi}(I_t,\tilde{L}_a),
	\end{equation}
	where $\phi$ is the parameter of SimNet, and  $\hat{W}$ is the predicted similarity map. 
	Each $\hat{W}(x)$ represents the similarity between $\tilde{L}_a$ and $I_t$ on spatial point $x$. 
	Meanwhile, by using entire images ($\tilde{L}_a$ and $I_t$) instead of local patches, SimNet is supposed to extract more discriminative latent features for similarity estimation. 
	
	The architecture of SimNet is presented in Fig. \ref{fig:framework} (c).
	The intermediate feature maps $(f_1, f_2, \dots, f_M)$ in the different levels of the network can capture the representative information of the input data with different scales. 
	In the output layer, SimNet upsamples $(f_1, f_2, \dots, f_M)$ to the same spatial size of $I_t$, and concatenates them to compute similarities. 
	Such network design can capture multi-scale information from images for fusion weights calculation.

	SimNet is trained by minimizing the cross entropy loss, which is defined as,
	\begin{equation}
		\label{eq:simloss}
		\mathcal{L}_{CE} = \\
		-\sum \limits_{x \in \Omega}^{} W_{gt}(x)log(\hat{W}(x))+(1-W_{gt}(x))log(1-\hat{W}(x)),
	\end{equation}
	where $W_{gt}$ is the ground truth similarity between $\tilde{L}_a$ and $I_t$.
	However, it is hard to obtain reliable $W_{gt}$ between label and intensity images for training. To resolve this, 
	we prepare $W_{gt}$ via the anatomical information from $\tilde{L}_a$ and $L_t$ (gold standard label of $I_t$).
	Suppose given a pair of $\tilde{L}_a$ and $L_t$, 
	each element of $W_{gt}$ is computed as follows,
	\begin{equation}
		\label{eq:gtsim}
		W_{gt}(x)= \frac{1}{|\mathcal{N}_x|}{\sum\limits_{v \in \mathcal{N}_x} \delta(\tilde{L}_a(v) \equiv L_t(v) )},
	\end{equation}     
	where $x \in \Omega$,  $\mathcal{N}_x$ is a local patch centering at $x$, and $|\mathcal{N}_x|$ is the size of $\mathcal{N}_x$. $W_{gt}(x)$ measures the probability of $\tilde{L}_a$ and $I_t$ having same anatomical structure on $x$.
	The predicted $\hat{W}$ can be directly applied as $W$ for cross-modality multi-atlas fusion in \eqref{equ:fusion}.   
	
	\section{Experiments and Results}
	\label{sec:exp}

	\subsection{{Datasets}}
	The framework was validated by two segmentation tasks, i.e, LV (including LVC and Myo), \hlo{ascending aorta (AA) and left atrium blood cavity (LAC)} segmentation on the MM-WHS dataset \cite{journal/mia/zhuang2019evaluation} and liver segmentation on the CHAOS dataset \cite{joural/mia/kavur2020chaos}. 
	\begin{itemize}
		\item the  MM-WHS dataset provides cardiac CT and MR images. 
		We adopted 40 (20 CT and 20 MR) images with corresponding manual labels to perform LV segmentation. 
		The CT images were acquired from two state-of-the-art CT scanners with a resolution of $0.78\times0.78\times1.60$ mm. 
		The MR images are balanced steady state free precession (b-SSFP) sequences acquired from a 1.5 T Philips scanner or a Siemens Magnetom Avanto 1.5 T scanner, and their resolutions are  $(1.6\sim2)\times(1.6\sim2)\times(2\sim3.2)$ mm. 
		\item the  CHAOS dataset provides multi-modality abdominal images. We adopted 40 (20 MR and 20 CT) images as well as their corresponding labels for liver segmentation.  
		The CT images were acquired from three state-of-the-art CT scanners with resolutions of $(0.7\sim0.8)\times(0.7\sim0.8)\times(3\sim3.2)$ mm$^3$. 
		The MR images are spectral pre-saturation inversion recovery (SPIR) sequences acquired from a 1.5 T Philips scanner, and their resolutions are $(1.36\sim1.89)\times(1.36\sim1.89)\times(5.5\sim9)$ mm$^3$.
	\end{itemize}

	For the image pre-processing, we re-sampled the original medical images to an isotropic resolution of $1 \times 1 \times 1$ mm$^3$, and then extracted sub-images from the task-relevant regions of interest. 
	\hlo{Next}, the sub-images were normalized to zero-mean and unit variance by z-score.
	\hlo{Finally, we performed rigid translation between a pair of atlas and target sub-images to align the geometric centers of the sub-images. Specifically, we employed the vector, which is formed from the center of the atlas sub-image to the center of the target sub-image, as the rigid translation parameter. 
	}
	
	\subsection{{Implementations}}

	For both MM-WHS and CHAOS datasets, the cross-modality MAS was conducted in two directions:  using CT atlases to segment MR targets (CT$\rightarrow$MR), or using MR atlases to segment CT targets (MR$\rightarrow$CT).
	We implemented our cross-modality MAS framework, including BiRegNet and SimNet, by TensorFlow on NVIDIA 1080 GPU. 
	
	To train BiRegNet, we fed the network with pairs of atlas and target image. By setting the balance weight $\lambda$ to the optimal $0.1$, we trained the network by minimizing the loss function (see \eqref{equ:reg_loss}) with a default Adam optimizer (learning rate $=$ 0.001, $\beta_1=0.9$, $\beta_2=0.999$).
	To train SimNet, we first registered the cross-modality atlases to the target image via BiRegNet, and achieved the corresponding warped atlases. Then, by setting the size of  $\mathcal{N}_x$ to the optimal $3\times 3\times 3$, we could prepare the ground truth similarities ($W_{gt}$) based on the warped atlas labels and the target gold standard labels (see \eqref{eq:gtsim}). Finally, we inputted pairs of warped atlas label and target image into the network, and optimized the network parameter by minimizing the loss function (see \eqref{eq:simloss}) via a default Adam optimizer. 
	\hlo{Besides, to avoid over-fitting,  data augmentation using random spatial transformation was employed.}
	
	To verify the registration methods, we calculated the Dice score (DS) and average symmetric surface distance (ASD) \cite{jouranl/mia/dou20173d}  between the warped atlas labels and the target gold standard labels. 
	To evaluate the performance of label fusion methods, we computed the DS, ASD, \hlo{volume difference (VD) and Hausdorff distance (HD)}  between the predicted target labels and the target gold standard labels. We employed Student T-test to decide whether differences are significant. 
	All results in our experiment were reported by cross-validation. 
	In each validation, we used 30 (15 CT and 15 MR) labelled images as cross-modality atlases, and the rest (5 CT and 5 MR) as target images. 
	Both BiRegNet and SimNet were trained by using the atlases.

	\subsection{{Results of Registration} }
		\begin{figure*}[htp]
	\centering
	\includegraphics[width=1\textwidth]{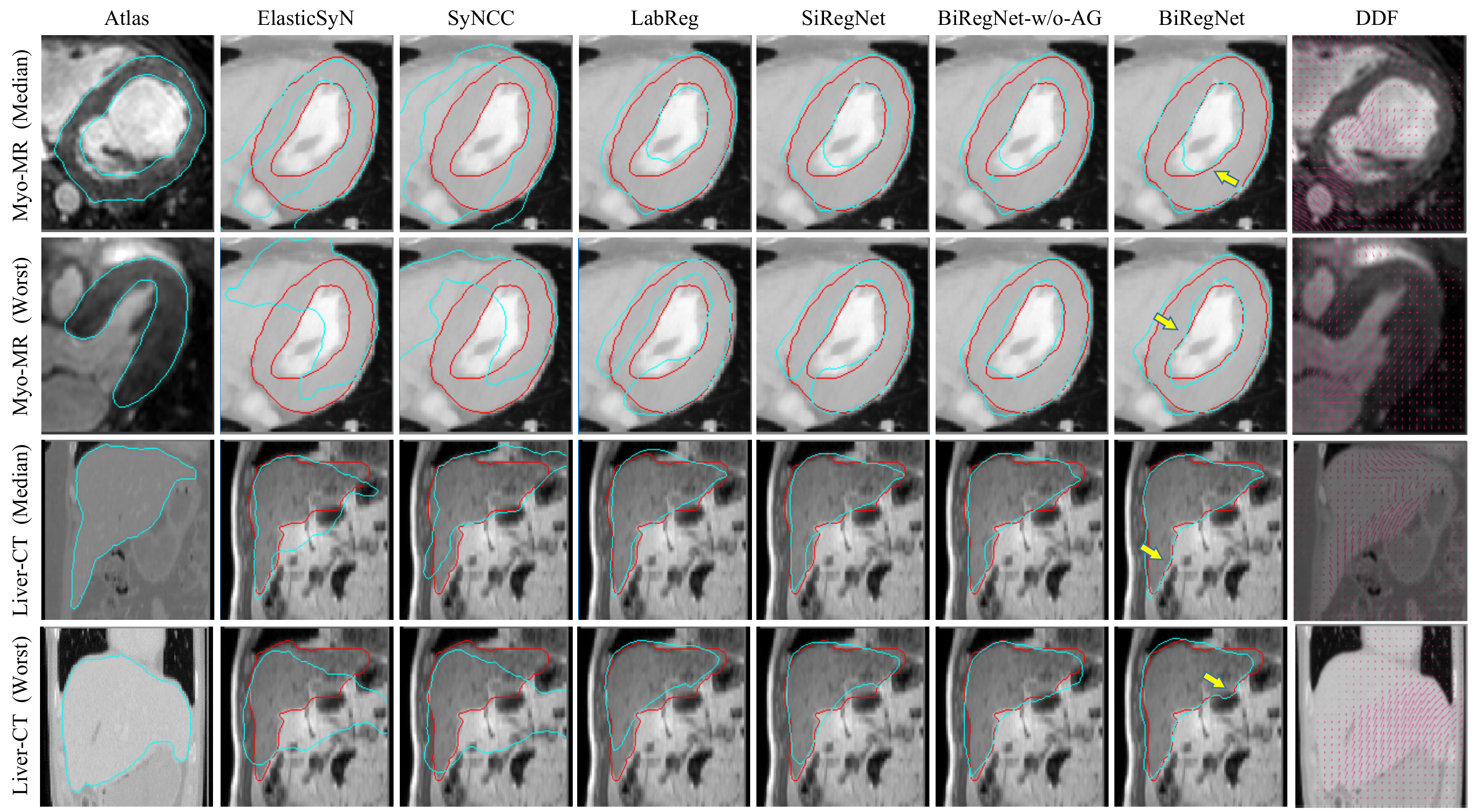}
	\caption{Visualization of Myo and liver registration results. The presented images are the median and worst cases in terms of Dice score of BiRegNet. The red contours delineate the gold standard labels, and the cyan contours delineate the (warped) atlas labels. The arrows point to the regions where BiRegNet achieves better visual results than the other registration methods. \hlo{The dense arrows in last column are used to visualize the DDF of BiRegNet. Additional registration results can be found in the Supplementary Material.} (The reader is referred to the colorful web version of this article.)   
	}
	\label{fig:reg}    
	\end{figure*}
	\label{sec:res:reg}
	To evaluate the performance of BiRegNet, we compared it with five registration methods: 
	\begin{itemize}
		\item SyNCC: The conventional affine + deformable registration, which is based on the symmetric image normalization method (SyN) with cross-correlation (CC) as optimization metric. 
		Note that the SyN is one of the top-performed registration methods \cite{journal/ni/klein2009evaluation}. 
		We implemented it based on the popular ANTs\footnote{\url{https://github.com/ANTsX/ANTsPy}} \cite{journal/inj/avants2009advanced} software package. 
		\item ElasticSyN: SyNCC registration method which use MI instead of CC as optimization metric.

		\item LabReg: The state-of-the-art cross-modality registration network \cite{journal/mia/hu2018weakly}. 
		We adopted their official implementation online\footnote{\url{https://github.com/YipengHu/label-reg}}.
		
		
		\item SiRegNet: BiRegNet without the consistency loss, i.e., a single-direction registration network. We compared this implementation to investigate the effect of the proposed consistency loss. 
		\item BiRegNet-W/o-AG: BiRegNet which utilizes channel wise concatenation instead of AG module for skip connections. We compared this implementation to study the effect of AG module. 
		\item BiRegNet: The proposed cross-modality registration network.
	\end{itemize}

	Table \ref{tab:reg_result} provides the quantitative results of the methods. 
	One can see that BiRegNet outperformed the conventional methods (SyNCC and ElasticSyN) on both datasets. 
	This is reasonable as BiRegNet took the advantages of anatomical labels to optimize the model parameter, which made it more reliable than the methods using intensity-based similarities (MI and CC).  
	Meanwhile, compared to the state-of-the-art cross-modality registration network (LabReg), BiRegNet achieved better results, \ie, DS and ASD, on the both datasets. It indicates that BiRegNet could achieve promising performance for cross-modality image registration.
	\begin{table*}[htp]
		\centering
		\caption{The performance of different registration methods. We compared the proposed method (BiRegNet) with three state-of-the art registration methods (SyNCC, ElasticSyN and LabReg), and two variants (SiRegNet and BiRegNet-W/o-AG) of BiRegNet. The best and second results are in bold and underlined, respectively.}
		\label{tab:reg_result}

		\resizebox{1\textwidth}{!}{ 
			
			\begin{tabular}{ll|llllllll|ll}
				\hline
				
				&\multicolumn{1}{l}{\multirow{2}{*}{Method}} & \multicolumn{2}{|c}{LVC}  & \multicolumn{2}{c}{Myo} & \multicolumn{2}{c}{\hlo{LAC}} & \multicolumn{2}{c}{\hlo{AA}}  & \multicolumn{2}{|c}{Liver}  \\ 
				\cline{3-12} 
				
				&	\multicolumn{1}{l}{}  
				& \multicolumn{1}{|c}{DS (\%)$\uparrow$}  & \multicolumn{1}{c}{ASD (mm)$\downarrow$} 
				& \multicolumn{1}{c}{DS (\%)$\uparrow$} &	\multicolumn{1}{c}{ASD (mm)$\downarrow$} 
				& \multicolumn{1}{c}{\hlo{DS (\%)$\uparrow$}} &	\multicolumn{1}{c}{\hlo{ASD (mm)$\downarrow$}} 
				& \multicolumn{1}{c}{\hlo{DS (\%)$\uparrow$}} &	\multicolumn{1}{c}{\hlo{ASD (mm)$\downarrow$}} 
				& \multicolumn{1}{|c}{DS (\%)$\uparrow$} & \multicolumn{1}{c}{ASD (mm)$\downarrow$} \\ 
				
				\hline
				
				\multirow{6}{*}{\rotatebox{90}{CT$\rightarrow$MR}}	&	\multicolumn{1}{l|}{SyNCC \cite{journal/inj/avants2009advanced}}  &70.79$\pm$16.89 & {4.59$\pm$3.37}  & 52.94$\pm$17.35 &    {3.42$\pm$1.50} 
				&\hlo{64.49$\pm$15.85} & \hlo{5.07$\pm$3.26} &\hlo{45.55$\pm$25.51} & \hlo{9.11$\pm$5.96}  &  \multicolumn{1}{|l}{75.13$\pm$12.48} & {8.56$\pm$6.15} \\
				
				&	\multicolumn{1}{l|}{ElasticSyN \cite{journal/inj/avants2009advanced}}  &69.79$\pm$16.65& {4.61$\pm$2.67}  & 53.89$\pm$17.24&    {3.50$\pm$1.48}
				&\hlo{67.32$\pm$13.25} & \hlo{4.47$\pm$2.28} &\hlo{46.38$\pm$24.78} & \hlo{8.53$\pm$5.92}  &  \multicolumn{1}{|l}{76.00$\pm$10.38} & {8.03$\pm$4.21} \\
				
				&	\multicolumn{1}{l|}{LabReg \cite{journal/mia/hu2018weakly}}  & {89.15$\pm$3.51}  & {1.77$\pm$0.45}  &  \underline{75.21$\pm$5.37}  & {1.97$\pm$0.5}  
				&\hlo{85.21$\pm$4.30} & \hlo{2.02$\pm$0.70} &\hlo{76.04$\pm$12.46} & \hlo{3.09$\pm$2.08}
				&\multicolumn{1}{|l}{87.58$\pm$1.77} & {4.05$\pm$0.58}\\
				
				
				&	\multicolumn{1}{l|}{SiRegNet} & \textbf{89.72$\pm$3.54}  & \textbf{1.67$\pm$0.42}  &  74.47$\pm$5.15  & {2.02$\pm$0.44}  
				&\hlo{85.38$\pm$4.22} & \hlo{1.98$\pm$0.66} &\hlo{78.80$\pm$11.90} & \hlo{\underline{2.73$\pm$1.89}}
				&  \multicolumn{1}{|l}{\underline{88.85$\pm$1.64}} & {\underline{3.61$\pm$0.49}}\\

				&	\multicolumn{1}{l|}{BiRegNet-W/o-AG}  & 88.85$\pm$4.01  & \multicolumn{1}{l}{1.81$\pm$0.46}  &  75.14$\pm$5.11  & \multicolumn{1}{l}{\underline{1.95$\pm$0.45}}  
				&\hlo{\underline{85.92$\pm$3.91}} & \hlo{\textbf{1.88$\pm$0.54}} &\hlo{\underline{79.07$\pm$11.27}} & \hlo{{2.74$\pm$1.90}}
				&  \multicolumn{1}{|l}{88.51$\pm$1.82} & {3.81$\pm$0.61}\\
				
				&		\multicolumn{1}{l|}{BiRegNet}    & \underline{89.23$\pm$4.19}  & \multicolumn{1}{l}{\underline{1.75$\pm$0.5}}  &  \textbf{76.46$\pm$5.06}  & \textbf{1.86$\pm$0.42}  
				&\hlo{\textbf{86.16$\pm$3.72}} & \hlo{\underline{1.89$\pm$0.65}} &\hlo{\textbf{80.87$\pm$9.66}} & \hlo{\textbf{2.50$\pm$1.69}}
				&  \multicolumn{1}{|l}{\textbf{89.47$\pm$1.65}} & \textbf{3.56$\pm$0.58}\\

				\hline
				\hline	
				
				\multirow{6}{*}{\rotatebox{90}{MR$\rightarrow$CT}}&	\multicolumn{1}{l|}{SyNCC \cite{journal/inj/avants2009advanced}}  &70.07$\pm$16.57 & {4.51$\pm$2.67}  & 50.66$\pm$16.02 &    {4.10$\pm$1.77}  
				&\hlo{67.43$\pm$14.62} & \hlo{4.66$\pm$2.14} 
				&\hlo{62.25$\pm$27.08} & \hlo{5.45$\pm$5.66}
				&\multicolumn{1}{|l}{74.49$\pm$9.53} & {8.75$\pm$3.40} \\
				
				&	\multicolumn{1}{l|}{ElasticSyN \cite{journal/inj/avants2009advanced}}  & 69.16$\pm$15.25 & {4.66$\pm$2.54}  & 49.00$\pm$16.21&    {4.34$\pm$2.04}  
				&\hlo{68.12$\pm$14.20} & \hlo{4.61$\pm$2.30} &
				\hlo{62.73$\pm$23.41} & 
				\hlo{4.82$\pm$3.26}
				&\multicolumn{1}{|l}{74.01$\pm$8.79} & {8.84$\pm$3.06} \\
				&	\multicolumn{1}{l|}{LabReg \cite{journal/mia/hu2018weakly}}  & 89.36$\pm$3.65  & \multicolumn{1}{l}{\underline{1.44$\pm$0.40}}  &  78.40$\pm$6.63  & {1.77$\pm$0.38}  
				&\hlo{88.93$\pm$3.20} & \hlo{1.61$\pm$0.42} &\hlo{87.16$\pm$11.17} & \hlo{1.62$\pm$1.27}
				&  \multicolumn{1}{|l}{87.78$\pm$3.39} & {4.12$\pm$1.22}\\
				
				
				&	\multicolumn{1}{l|}{SiRegNet} & \underline{89.38$\pm$3.69}  & {1.44$\pm$0.44}  &  \underline{78.89$\pm$5.71}  & {\underline{1.73$\pm$0.31}}  
				&\hlo{89.55$\pm$3.82} & \hlo{1.49$\pm$0.44} & \hlo{87.04$\pm$12.82} & \hlo{1.68$\pm$1.60}
				&  \multicolumn{1}{|l}{\underline{88.6$\pm$3.90}} & {\textbf{3.84$\pm$1.50}}\\

				&	\multicolumn{1}{l|}{BiRegNet-W/o-AG}  & 88.17$\pm$3.86  & {1.58$\pm$0.35}  &  77.77$\pm$5.01  & {1.85$\pm$0.28}  
				&\hlo{\underline{89.99$\pm$3.11}} & \hlo{\underline{1.44$\pm$0.34}} &\hlo{\underline{88.01$\pm$8.21}} & \hlo{\underline{1.59$\pm$1.15}}
				&  \multicolumn{1}{|l}{87.78$\pm$4.64} & {4.21$\pm$1.64}\\
				

				&		\multicolumn{1}{l|}{BiRegNet}   & \textbf{90.24$\pm$3.56}  & \textbf{1.30$\pm$0.32}  &  \textbf{80.39$\pm$4.49}  & \textbf{1.66$\pm$0.33} 
				&\hlo{\textbf{90.23$\pm$3.39}} & \hlo{\textbf{1.39$\pm$0.35}} &\hlo{\textbf{88.48$\pm$14.98}} & \hlo{\textbf{1.52$\pm$1.63}}
				&  \multicolumn{1}{|l}{\textbf{88.54$\pm$5.09}} & \underline{3.91$\pm$1.71}\\
				
				\hline
				
			\end{tabular}
		}

	\end{table*}

\begin{figure}[htp]
		\centering
		\includegraphics[width=\columnwidth]{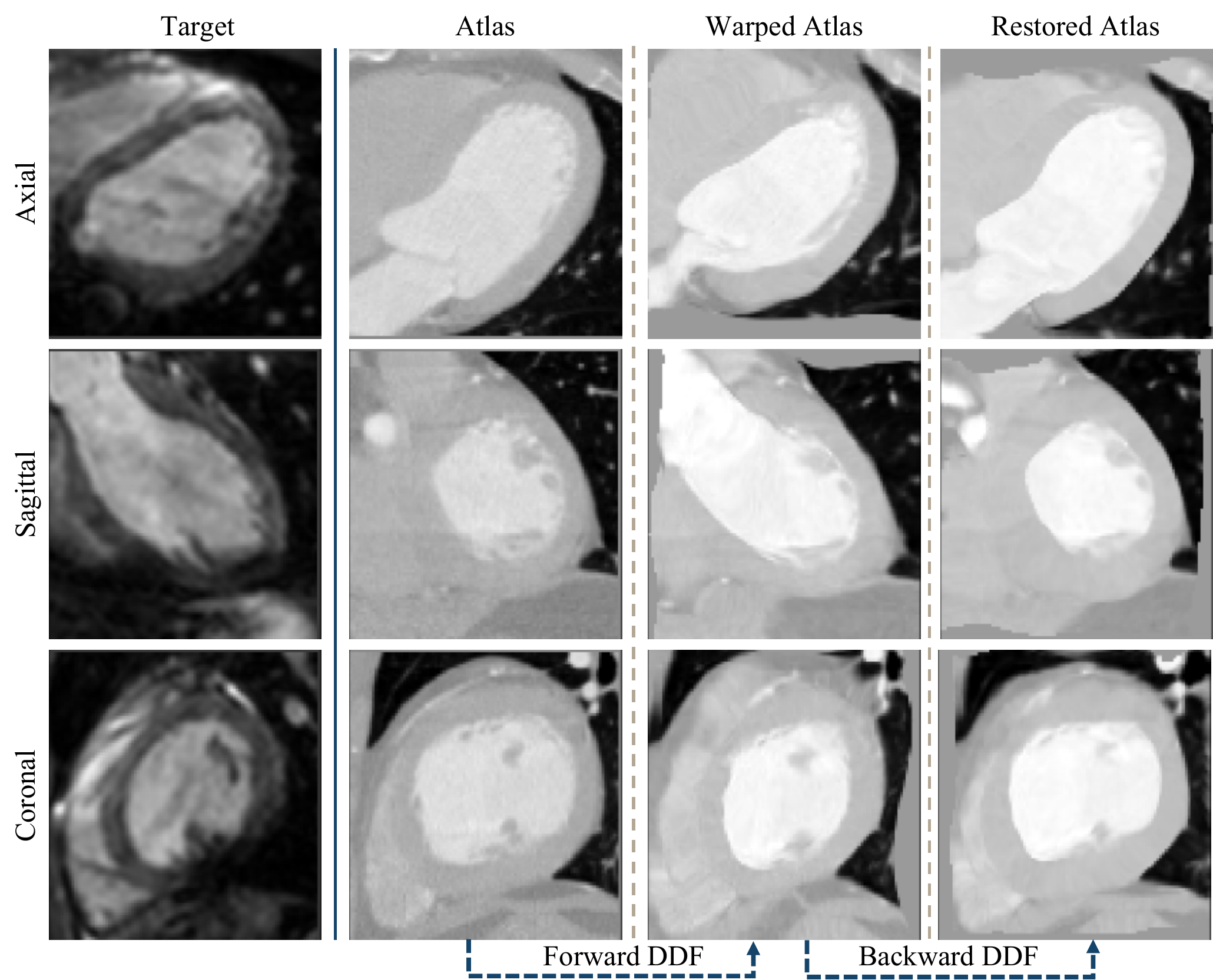}
		\caption{Visualization of a restoration procedure. 
			We first estimated a pair of forward and backward DDFs between the atlas image and the target image via BiRegNet. 
			Then, we transformed the atlas via the forward DDF to obtain the warped atlas, and transformed the warped atlas via the backward DDF for the restored atlas. 
			The slices were extracted from the same spatial position along three different axes (axial, sagittal, and coronal). }
		\label{fig:restore}    
	\end{figure}

	SiRegNet and BiRegNet achieved comparable result for the LVC and Liver registration. \hlo{For LAC and AA registration, BiRegNet could obtain slightly better results than SiRegNet.} For the most challenging Myo registration task, BiRegNet obtained average 1.7\% (CT$\rightarrow$MR: 76.46\% VS. 74.47\%, $p<0.001$; MR$\rightarrow$CT: 80.39\% VS. 78.89\%, $p<0.001$) and 0.11 mm (CT$\rightarrow$MR: 1.86 mm VS. 2.02 mm, $p<0.001$; MR$\rightarrow$CT: 1.66 mm VS. 1.73mm, $p<0.001$) improvements on DS and ASD metrics against SiRegNet, respectively. This demonstrates the advantage of consistency loss for regularizing the registration network. Additionally, Fig. \ref{fig:restore} investigates the consistency of forward and backward DDFs. We transformed an atlas by using a pair of forward and backward DDFs successively to obtain a restored atlas. Ideally, the restored atlas should be equal to the original one when the forward and backward DDFs define consistent correspondences. As expected, the restored atlas slices were visually similar to their original ones. It implies that the consistency loss could force BiRegNet to generate consistent forward and backward DDFs. 
	
\begin{figure}[htp]
		\centering		
		\includegraphics[width=0.85\columnwidth]{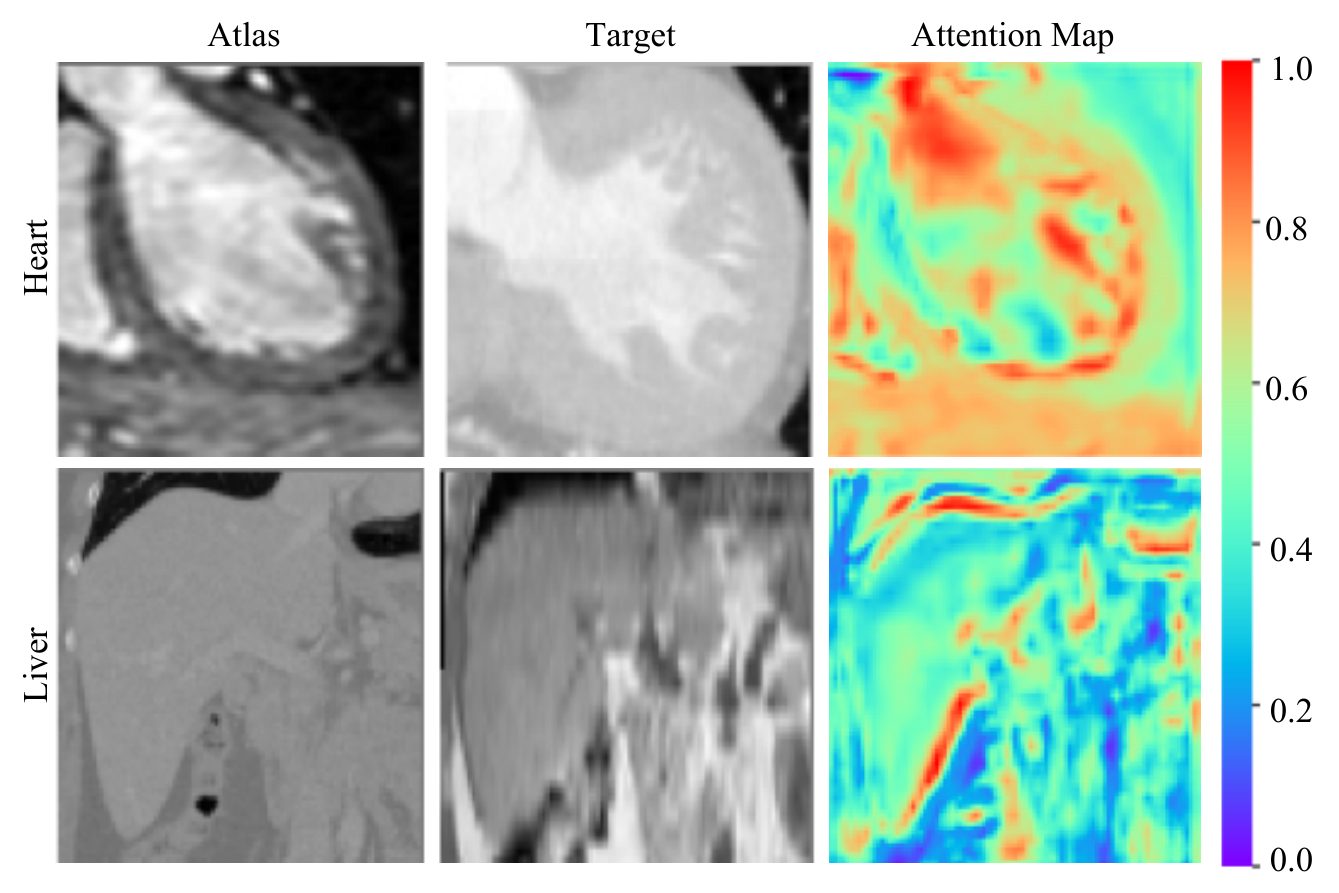}
		\caption{Visualization of AG coefficient. 
			The showed attention maps are from the last AG module of BiRegNet (see Fig. \ref{fig:framework} (b)). 
			The activation maps are consistently corresponded to the perceptually misaligned areas between atlas and target images.}
		\label{fig:AG_vis}    
	\end{figure}

	Moreover, without AG module, BiRegNet-W/o-AG suffered from the decline of performance on \hlo{LVC, Myo, AA and liver regions}. For example, on CHAOS dataset, the average DS dropped from 89.00\% (CT$\rightarrow$MR: 89.47\%, MR$\rightarrow$CT 88.54\%) to 88.14\% (CT$\rightarrow$MR: 88.51\%, MR$\rightarrow$CT 87.78\%), and the average ASD increased from 3.73 mm (CT$\rightarrow$MR: 3.56 mm, MR$\rightarrow$CT 3.91 mm) to 4.01 mm (CT$\rightarrow$MR: 3.81 mm, MR$\rightarrow$CT 4.21 mm). It proves the effectiveness of AG module. Additionally, Fig. \ref{fig:AG_vis} visualizes two representative attention coefficients $\alpha_{att}$ (see \eqref{eq:att}) for a deep understanding of AG module. The coefficients aim to re-calibrate propagated feature maps to extract several specific information for the registration. For both cases, one can observe that  $\alpha_{att}$ emphasized the areas which are not initially aligned between the atlas and the target images. 
	This is reasonable because such misaligned areas require larger transformation for registration.

	Fig. \ref{fig:reg} presents four representative registration cases from the MM-WHS and CHAOS dataset. It shows that the DNN-based registration methods (LabReg, BiRegNet, SiRegNet and BiRegNet-W/o-AG) achieved more reasonable visual results than the conventional methods (SyNCC and ElasticSyN), which is consistent with the quantitative results in Table \ref{tab:reg_result}. 
	Note that although BiRegNet only showed slight improvements in terms of DS and ASD when compared to other DNN-based registration methods, it could capture more reasonable details as indicated by the arrows in Fig. \ref{fig:reg}. \hlo{In addition, the last column of Fig. \ref{fig:reg} shows the DDFs of BiRegNet overlapped on atlas images. The directions and lengths of arrows indicate the directions and magnitudes of displacement vectors, respectively.}
	
    \hlo{Furthermore, Table \ref{tab:parameter:time} (a) presents the computational load. All learning-based registration methods (LabReg, SiRegNet, BiRegNet-W/o-AG and BiRegNet) required a large memory to load the DNN models, while their average run time for registration was promising, \ie, within 2 seconds. In contrast, an iterative conventional method, such as SyNCC, cost 106.7 seconds to perform registration. 
    }

	\begin{table}[htp]
		
		\centering
		\caption{\coloro{The size of parameters and average run time of different registration and label fusion methods. Note that the number of parameters is only reported  for learning-based registration (LabReg, SiRegNet, BiRegNet-W/o-AG and BiRegNet) and label fusion (MAS-Patch and MAS-SimNet) methods.}}

			\resizebox{\columnwidth}{!}{
		    			\begin{tabular}{l|cccccc}
				\hline
					\multicolumn{7}{c}{\hlo{(a) Registration}} \\
				\hline
				 & \hlo{SyNCC} & \hlo{ElasticSyN} & \hlo{LabReg} & \hlo{SiRegNet}   & \hlo{BiRegNet-W/o-AG} & \hlo{BiRegNet} \\
				\hline
				\hlo{Parameters}  & \hlo{N/A}   & \hlo{N/A}   & \hlo{307k}   &  \hlo{239k} & \hlo{248K} & \hlo{250k} \\
				\hline
				\hlo{Run Time (s)} &  \hlo{106.7}  & \hlo{27.7}   & \hlo{1.3}  & \hlo{0.5}   & \hlo{0.9} & \hlo{0.9} \\

				\hline
					\multicolumn{7}{c}{\hlo{(b) Label Fusion}} \\
				\hline
				 & MV & STAPLE & LWF-MI & JLF\hlo{$_{ei}$}   & \hlo{MAS-Patch} & MAS-SimNet \\
				\hline 
				\hlo{{Parameters}}  & \hlo{N/A}   & \hlo{N/A}   & \hlo{N/A}   &   \hlo{N/A} & \hlo{18k} &  \hlo{4.7M} \\
				\hline
				Run Time (s) & 0.9   & 1.3   & 469.1   &   479.8 & {\hlo{353.7}} & 8.3 \\
                \hline
			\end{tabular}%
				}

		\label{tab:parameter:time}%
	\end{table}%

	\subsection{{Results of Label Fusion}}
	
	\subsubsection{Comparisons with Different Label Fusion Methods}
	
	\label{sec:res:lf}
	
	To analyze the performance of the proposed fusion method, we compared it with four potential cross-modality multi-atlas fusion methods. Note that the performances of all fusion methods were reported by fusing the warped atlases obtained from BiRegNet.
	\begin{itemize}
		\item MV:  Majority voting label fusion strategy. 
		This method can be directly employed for cross-modality fusion because it assumes all the atlases have the same fusion weights.
		\item STAPLE: The simultaneous truth and performance level estimation label fusion strategy  \cite{journal/tmi/warfield2004simultaneous}. 
		This method is suitable for cross-modality fusion as it estimates the probability of the true segmentation only using the warped atlas labels. We implemented it based on SimpleITK\footnote{https://github.com/SimpleITK/SimpleITK}. 
		\item LWF-MI: LWF strategy \cite{journal/tmi/artaechevarria2009combination}, where the MI between the warped atlas image patches and the target image patch were employed as the fusion weights. 
		To calculate the MI metric, the optimal size of the local patch was set to $7 \times 7 \times 7$. 
		\item JLF\hlo{$_{ei}$}: The state-of-the-art joint label fusion strategy \cite{jour/pami/wang2012multi}. The patch size and search range of JLF\hlo{$_{ei}$} were set to $7 \times 7 \times 7$ and $3\times 3 \times 3$, respectively. Meanwhile, the intensity of patches were normalized by z-score when calculating the pairwise dependency matrix of JLF\hlo{$_{ei}$} \hlo{based on entropy images \cite{journal/mia/wachinger2012entropy}}. We employed the implementation online\footnote{http://www.sdspeople.fudan.edu.cn/zhuangxiahai/}.  
		\item \hlo{MAS-Patch \cite{conf/miccai/ding2020cross}: The learning-based cross-modality multi-atlas fusion method, where the fusion weights were calculated based on local intensity image patches. The patch size was set to $15 \times 15 \times 15$.}
		\item MAS-SimNet: Our proposed cross-modality multi-atlas fusion method, where the fusion weights between the warped atlases and the target image are estimated by SimNet. 
		
	\end{itemize}
	
	
	 \hlo{The top part of Table \ref{tab:mas} summarizes the DS and ASD results of different label fusion methods.} MAS-SimNet exhibited better DS (LVC: $p<0.001$, Myo: $p<0.001$, \hlo{LAC: $p<0.001$, AA: $p=0.009$}, Liver: $p<0.005$) and ASD (LVC: $p<0.001$, Myo: $p<0.001$, \hlo{LAC: $p<0.001$, AA: $p=0.023$}, Liver: $p<0.001$) results than MV on both datasets. \hlo{Meanwhile, except for the segmentation of AA, MAS-SimNet could obtain better DS and ASD than STAPLE for all structures.} Compared to LWF-MI, MAS-SimNet especially achieved better average DS (CT$\rightarrow$MR: 81.06\% VS. 79.50\%, $p<0.001$; MR$\rightarrow$CT: 85.71\%  VS. 83.77\%, $p<0.001$) and ASD (CT$\rightarrow$MR: 1.53 mm VS. 1.65mm, $p<0.001$; MR$\rightarrow$CT: 1.28 mm VS. 1.46 mm, $p<0.001$) for the most challenging Myo segmentation. \hlo{In addition, the bottom part of Table \ref{tab:mas} also presents the VD and HD results of different label fusion methods. Overall, MAS-SimNet had better VD and HD results than MV, STAPLE and LWF-MI on both datasets. This proved that MAS-SimNet could achieve state-of-the-art performances for different anatomy structures.}
	
   	\hlo{In Table \ref{tab:mas}, one can also observe the top performances in terms of DS, ASD, VD and HD of different anatomy structures (LVC, Myo, LAC, AA and Liver) were achieved by JLF\hlo{$_{ei}$}, MAS-Patch or MAS-SimNet. Notably, except for the quantitative results in Table \ref{tab:mas}, the computational load for label fusion methods needs to be taken into consideration \cite{journal/mia/iglesias2015multi}. Table \ref{tab:parameter:time} (b) presents the computational load of each method.} MV, STAPLE and MAS-SimNet were computationally efficient. They could perform a label fusion step within seconds, whereas LWF-MI, JLF\hlo{$_{ei}$} and \hlo{MAS-Patch} required more than \hlo{350 seconds}. The reason is that LWF-MI, JLF\hlo{$_{ei}$} and \hlo{MAS-Patch} were designed to calculate fusion weights of atlases patch-wisely, while MAS-SimNet estimates fusion weights of atlases by using DNN in one-step fashion. \hlo{Here, patch-based fusion weight estimation is time consuming. For instance, as listed in Table \ref{tab:parameter:time} (b), although MAS-SimNet had more parameters than MAS-Patch (4.7M VS 18K), it yielded almost 40 times faster (8.3 seconds VS 353.7 seconds). MAS-SimNet could obtain comparable performances to JLF\hlo{$_{ei}$} and MAS-Patch in a more computation-efficient way.}        
	
\begin{table*}[htp]

		\centering
		\caption{Comparison of segmentation performance between MAS-SimNet and other state-of-the-art label fusion methods. The best and second results are in bold and underlined, respectively.}
		\label{tab:mas}
		    		
		\resizebox{1\textwidth}{!}{
			
			\begin{tabular}{ll|llllllll|ll}
				\hline
				
				&	\multicolumn{1}{l}{\multirow{2}{*}{Method}} & \multicolumn{2}{|c}{LVC}  & \multicolumn{2}{c}{Myo}   &\multicolumn{2}{c}{\hlo{LAC}}  & \multicolumn{2}{c}{\hlo{AA}}   & \multicolumn{2}{|c}{Liver}  \\ 
				\cline{3-12} 
				
				&	\multicolumn{1}{l}{}  
				& \multicolumn{1}{|c}{DS (\%)$\uparrow$}  & \multicolumn{1}{c}{ASD (mm)$\downarrow$} 
				& \multicolumn{1}{c}{DS (\%)$\uparrow$} &	\multicolumn{1}{c}{ASD (mm)$\downarrow$} 
				& \multicolumn{1}{c}{\hlo{DS (\%)$\uparrow$}} &	\multicolumn{1}{c}{\hlo{ASD (mm)$\downarrow$}} 
				& \multicolumn{1}{c}{\hlo{DS (\%)$\uparrow$}} &	\multicolumn{1}{c}{\hlo{ASD (mm)$\downarrow$}} 
				& \multicolumn{1}{|c}{{DS (\%)$\uparrow$}} & \multicolumn{1}{c}{{ASD (mm)$\downarrow$}} \\ 
				
				\hline
					{\multirow{6}{*}{\rotatebox{90}{CT$\rightarrow$MR}}}	&	\multicolumn{1}{l|}{MV}  & {90.05$\pm$4.01}  & {1.63$\pm$0.49}  &  79.39$\pm$4.49  & {1.67$\pm$0.35}  
				&\hlo{86.97$\pm$3.52} & \hlo{1.80$\pm$0.62} & \hlo{82.50$\pm$9.00} & \hlo{2.26$\pm$1.61}
				&  \multicolumn{1}{|l}{{90.57$\pm$1.43}} & {3.17$\pm$0.46}\\
				&	\multicolumn{1}{l|}{STAPLE \cite{journal/tmi/warfield2004simultaneous}}   & {89.66$\pm$4.22}  & {1.70$\pm$0.52}  &  78.26$\pm$4.47  & {1.85$\pm$0.35}  
				&\hlo{86.82$\pm$3.78} & \hlo{1.83$\pm$0.63} & \hlo{81.45$\pm$9.73} & \hlo{2.46$\pm$1.75}
				&  \multicolumn{1}{|l}{90.35$\pm$1.56} & {3.27$\pm$0.51}\\
				
				&	\multicolumn{1}{l|}{LWF-MI \cite{journal/tmi/artaechevarria2009combination}}  & {90.07$\pm$4.02}  & {1.63$\pm$0.50}  &  79.50$\pm$4.49  & {1.65$\pm$0.35}  
				&\hlo{86.98$\pm$3.54} & \hlo{1.80$\pm$0.63} &\hlo{82.62$\pm$8.84} & \hlo{2.29$\pm$1.64}
				&  \multicolumn{1}{|l}{\underline{90.71$\pm$1.42}} & {3.11$\pm$0.45}\\
				
				&	\multicolumn{1}{l|}{JLF\hlo{$_{ei}$} \cite{jour/pami/wang2012multi}} & {\textbf{91.25$\pm$3.68}}  & \textbf{1.44$\pm$0.51}  &  \textbf{81.38$\pm$4.20}  & \textbf{1.47$\pm$0.36}  
				&\hlo{\underline{87.57$\pm$3.87}} & \hlo{\underline{1.70$\pm$0.63}} &\hlo{\underline{83.60$\pm$8.00}} & \hlo{2.12$\pm$1.52}
				&  \multicolumn{1}{|l}{89.92$\pm$1.21} & {3.31$\pm$0.41}\\
				
				&	\multicolumn{1}{l|}{\hlo{MAS-Patch \cite{conf/miccai/ding2020cross}}} 
				&  \hlo{89.95$\pm$3.94} & \hlo{1.70$\pm$0.50} 
				& \hlo{78.99$\pm$4.52} & \hlo{1.54$\pm$0.30}  
				&\hlo{87.05$\pm$3.02}& \hlo{2.08$\pm$0.45} &\hlo{83.18$\pm$7.70} & \hlo{\underline{2.11$\pm$0.82}}
				&  \multicolumn{1}{|l}{\hlo{90.46$\pm$1.36}} & \hlo{\textbf{1.57$\pm$0.22}}\\

				&	\multicolumn{1}{l|}{MAS-SimNet} & \underline{90.78$\pm$3.56}  & \underline{1.52$\pm$0.43}  &  \underline{81.06$\pm$3.89}  & \underline{1.53$\pm$0.33}  
				&\hlo{\textbf{87.65$\pm$3.43}} & \hlo{\textbf{1.70$\pm$0.60}} &\hlo{\textbf{84.90$\pm$6.90}} & \hlo{\textbf{1.93$\pm$1.26}}
				&  \multicolumn{1}{|l}{\textbf{91.24$\pm$1.34}} & {\underline{2.92$\pm$0.41}}\\
				\hline
				\hline
				
				{\multirow{6}{*}{\rotatebox{90}{MR$\rightarrow$CT}}} &	\multicolumn{1}{l|}{MV} & {91.19$\pm$3.29}  & {1.18$\pm$0.26}  &  83.61$\pm$3.93  & {1.48$\pm$0.32}  
				&\hlo{91.39$\pm$3.27} & \hlo{1.25$\pm$0.34} &\hlo{89.51$\pm$16.66} & \hlo{1.41$\pm$2.01}
				&  \multicolumn{1}{|l}{89.96$\pm$4.77} & {3.51$\pm$1.76}\\
				&	\multicolumn{1}{l|}{STAPLE \cite{journal/tmi/warfield2004simultaneous}}  & {91.08$\pm$3.23}  & {1.21$\pm$0.27}  &  82.17$\pm$4.44  & {1.68$\pm$0.32}  
				&\hlo{91.41$\pm$3.05} & \hlo{1.25$\pm$0.31} &\hlo{\underline{90.62$\pm$10.50}} & \hlo{\underline{1.26$\pm$1.37}}
				&  \multicolumn{1}{|l}{90.31$\pm$3.76} & {3.40$\pm$1.43}\\
				
				&	\multicolumn{1}{l|}{LWF-MI \cite{journal/tmi/artaechevarria2009combination}}  & {91.24$\pm$3.29}  & {1.17$\pm$0.26}  &  83.77$\pm$3.86  & {1.46$\pm$0.32}  
				&\hlo{91.43$\pm$3.26} & \hlo{1.24$\pm$0.34} &\hlo{89.54$\pm$16.74} & \hlo{1.41$\pm$1.97}
				&  \multicolumn{1}{|l}{90.07$\pm$4.76} & {3.47$\pm$1.77}\\
				
				&	\multicolumn{1}{l|}{JLF\hlo{$_{ei}$} \cite{jour/pami/wang2012multi}} & \textbf{92.50$\pm$3.50}  & \textbf{0.97$\pm$0.27}  &  \underline{84.90$\pm$3.21}  & \underline{1.37$\pm$0.33}  
				&\hlo{\underline{91.83$\pm$3.09}} & \hlo{\underline{1.18$\pm$0.29}} &\hlo{90.54$\pm$12.29} & \hlo{\textbf{1.24$\pm$1.42}}
				&  \multicolumn{1}{|l}{\underline{90.39$\pm$4.19}} & {3.32$\pm$1.47}\\
				
				&	\multicolumn{1}{l|}{\hlo{MAS-Patch \cite{conf/miccai/ding2020cross}}} & \hlo{90.09$\pm$3.32}  & \hlo{1.44$\pm$0.34}  &  \hlo{83.12$\pm$4.10}  & \hlo{1.54$\pm$0.30}
				& \hlo{91.20$\pm$3.07} & \hlo{1.62$\pm$0.45} &\hlo{\textbf{90.95$\pm$10.57}} & \hlo{1.48$\pm$1.53}
				&  \multicolumn{1}{|l}{\hlo{89.91$\pm$3.86}} & \hlo{\textbf{1.74$\pm$0.73}}\\
				
				&	\multicolumn{1}{l|}{MAS-SimNet} & \underline{91.93$\pm$2.94}  & \underline{1.09$\pm$0.23}  &  \textbf{85.71$\pm$3.70}  & {\textbf{1.28$\pm$0.24}}  
				&\hlo{\textbf{92.30$\pm$3.15}} &\hlo{ \textbf{1.11$\pm$0.34}} &\hlo{89.70$\pm$19.53} & \hlo{1.42$\pm$2.42}	&\multicolumn{1}{|l}{\textbf{92.23$\pm$2.31}} & \underline{2.72$\pm$0.93}\\
				
				\hline \\
				
			
	            \hline
				&	\multicolumn{1}{l}{}  
				& \multicolumn{1}{|c}{\hlo{VD (mL)$\downarrow$}}  & \multicolumn{1}{c}{\hlo{HD (mm)$\downarrow$}} 	& \multicolumn{1}{c}{\hlo{VD (mL)$\downarrow$}}  & \multicolumn{1}{c}{\hlo{HD (mm)$\downarrow$}} 
				& \multicolumn{1}{c}{\hlo{VD (mL)$\downarrow$}} &	\multicolumn{1}{c}{\hlo{HD (mm)$\downarrow$}} & \multicolumn{1}{c}{\hlo{VD (mL)$\downarrow$}}  & \multicolumn{1}{c|}{\hlo{HD (mm)$\downarrow$}}
			
				& \multicolumn{1}{|c}{\hlo{VD (mL)$\downarrow$}} & \multicolumn{1}{c}{\hlo{HD (mm)$\downarrow$}} \\ 
				
				\hline
					{\multirow{6}{*}{\rotatebox{90}{\hlo{CT$\rightarrow$MR}}}}	&	\multicolumn{1}{l|}{\hlo{MV}}   & \hlo{12.85$\pm$11.29} & \hlo{10.90$\pm$3.73}   & \hlo{25.85$\pm$12.47} & \hlo{13.52$\pm$3.52} & \hlo{8.95$\pm$10.32} & \hlo{14.95$\pm$8.81} & \hlo{12.95$\pm$13.21} & \hlo{17.36$\pm$11.57}
				&  \hlo{128.57$\pm$81.80} & \hlo{23.68$\pm$4.78} \\
				&	\multicolumn{1}{l|}{\hlo{STAPLE \cite{journal/tmi/warfield2004simultaneous}}}   & \hlo{14.52$\pm$12.96} & \hlo{11.08$\pm$3.85}   & \hlo{40.61$\pm$16.62} & \hlo{13.63$\pm$3.26} & \hlo{8.68$\pm$9.93} & \hlo{14.87$\pm$8.87} & \hlo{19.97$\pm$22.34} & \hlo{18.63$\pm$12.84}
				&  \hlo{157.38$\pm$93.86} & \hlo{24.23$\pm$5.17} \\
				
				&	\multicolumn{1}{l|}{\hlo{LWF-MI \cite{journal/tmi/artaechevarria2009combination}}}  & \hlo{12.84$\pm$11.40} & \hlo{10.95$\pm$3.92}    & \hlo{25.12$\pm$12.10} & \hlo{13.61$\pm$3.42} & \hlo{8.92$\pm$10.35} & \hlo{15.03$\pm$8.83} & \hlo{12.54$\pm$12.65} & \hlo{16.60$\pm$11.45}
				&   \hlo{\underline{123.81$\pm$31.04}} & \hlo{23.62$\pm$4.82} \\
				
				&	\multicolumn{1}{l|}{\hlo{JLF\hlo{$_{ei}$} \cite{jour/pami/wang2012multi}}} & \hlo{\textbf{10.30$\pm$8.24}} & \hlo{11.38$\pm$4.02}    & \hlo{\textbf{12.02$\pm$10.19}} & \hlo{14.35$\pm$4.30} & \hlo{\textbf{7.99$\pm$10.28}} & \hlo{\underline{14.54$\pm$8.50}} & \hlo{12.35$\pm$12.76} & \hlo{17.19$\pm$11.68}
				&  \hlo{164.80$\pm$103.75} & \hlo{23.88$\pm$5.22} \\
				
				&	\multicolumn{1}{l|}{\hlo{MAS-Patch \cite{conf/miccai/ding2020cross}}} 
			& \hlo{12.90$\pm$10.95} & \hlo{\underline{10.64$\pm$3.68}} 
				 & \hlo{26.77$\pm$12.68} &\hlo{\underline{13.09$\pm$3.66}}& \hlo{8.93$\pm$10.31} & \hlo{\textbf{14.50$\pm$9.07}} & \hlo{\underline{7.36$\pm$6.21}} &\hlo{\textbf{14.76$\pm$9.82}}
				& \hlo{130.80$\pm$77.93} & \hlo{\textbf{22.14$\pm$4.32}}\\

				&	\multicolumn{1}{l|}{\hlo{MAS-SimNet}}  & \hlo{\underline{11.18$\pm$9.98}} & \hlo{\textbf{10.54$\pm$4.17}}   & \hlo{\underline{21.40$\pm$10.63}} & \hlo{\textbf{12.75$\pm$3.90}} & \hlo{\underline{8.43$\pm$10.27}} & \hlo{14.55$\pm$8.88} & \hlo{\textbf{7.31$\pm$6.07}} &\hlo{\underline{15.78$\pm$10.48}}
				&  \hlo{\textbf{112.48$\pm$71.44}} & \hlo{\underline{22.35$\pm$4.74}} \\
				\hline
				\hline
				
				{\multirow{6}{*}{\rotatebox{90}{\hlo{MR$\rightarrow$CT}}}} &	\multicolumn{1}{l|}{\hlo{MV}}& \hlo{5.25$\pm$3.22} & \hlo{6.78$\pm$1.97}   & \hlo{{15.06$\pm$8.78}} & \hlo{8.23$\pm$1.85} & \hlo{5.11$\pm$3.38} & \hlo{7.51$\pm$2.51} & \hlo{4.53$\pm$7.74} & \hlo{11.50$\pm$8.19}
				&   \hlo{189.91$\pm$176.31} & \hlo{28.72$\pm$10.90} \\
				&	\multicolumn{1}{l|}{\hlo{STAPLE \cite{journal/tmi/warfield2004simultaneous}}} &\hlo{5.70$\pm$3.85} & \hlo{\underline{6.68$\pm$1.89}}   & \hlo{25.13$\pm$13.44} & \hlo{\underline{8.09$\pm$1.17}}   & \hlo{5.60$\pm$2.71} & \hlo{7.54$\pm$2.46} & \hlo{6.53$\pm$12.77} & \hlo{11.48$\pm$8.19}
				&  \hlo{\underline{163.68$\pm$137.75}} & \hlo{28.16$\pm$9.87} \\
				
				&	\multicolumn{1}{l|}{\hlo{LWF-MI \cite{journal/tmi/artaechevarria2009combination}}} & \hlo{5.19$\pm$3.20} & \hlo{6.74$\pm$1.99}   & \hlo{\underline{14.93$\pm$8.59}} & \hlo{8.18$\pm$1.86}  & \hlo{5.13$\pm$3.35} & \hlo{7.53$\pm$2.52} & \hlo{4.51$\pm$7.69} & \hlo{11.62$\pm$8.25}
				&  \hlo{178.91$\pm$173.86} & \hlo{28.51$\pm$11.08} \\
				
				&	\multicolumn{1}{l|}{\hlo{JLF\hlo{$_{ei}$} \cite{jour/pami/wang2012multi}}}  & \hlo{\textbf{4.50$\pm$3.22}} & \hlo{6.98$\pm$1.58}   & \hlo{22.09$\pm$14.28} & \hlo{10.92$\pm$2.67}  & \hlo{6.19$\pm$2.95} & \hlo{8.25$\pm$2.59} & \hlo{6.19$\pm$12.41} & \hlo{\textbf{9.21$\pm$7.46}}
				&  \hlo{168.95$\pm$146.17} & \hlo{28.20$\pm$9.48} \\
				
				&	\multicolumn{1}{l|}{\hlo{MAS-Patch \cite{conf/miccai/ding2020cross}}} & \hlo{5.48$\pm$3.41} & \hlo{6.71$\pm$1.84} & \hlo{15.21$\pm$9.28} & \hlo{8.34$\pm$1.81} & \hlo{\underline{5.01$\pm$3.55}} & \hlo{\textbf{6.73$\pm$2.05}} & \hlo{\textbf{2.64$\pm$2.21}} & \hlo{9.62$\pm$7.99}
				& \hlo{178.05$\pm$169.58} & \hlo{\underline{25.92$\pm$8.07}} \\
				
			&	\multicolumn{1}{l|}{\hlo{MAS-SimNet}}  & \hlo{\underline{4.83$\pm$3.13}} & \hlo{\textbf{6.08$\pm$1.66}}   & \hlo{\textbf{13.22$\pm$8.86}} & \hlo{\textbf{7.66$\pm$2.10}} & \hlo{\textbf{4.51$\pm$3.04}} & \hlo{\underline{6.92$\pm$2.57}} & \hlo{\underline{3.57$\pm$5.22}} & \hlo{\underline{9.30$\pm$8.25}}	& \hlo{\textbf{129.38$\pm$ 111.09}} & \hlo{\textbf{23.80$\pm$6.94}} \\
				
				\hline
				
			\end{tabular}
		
		}

	\end{table*}

\begin{figure*}[htp]
		\centering
		\includegraphics[width=1\textwidth]{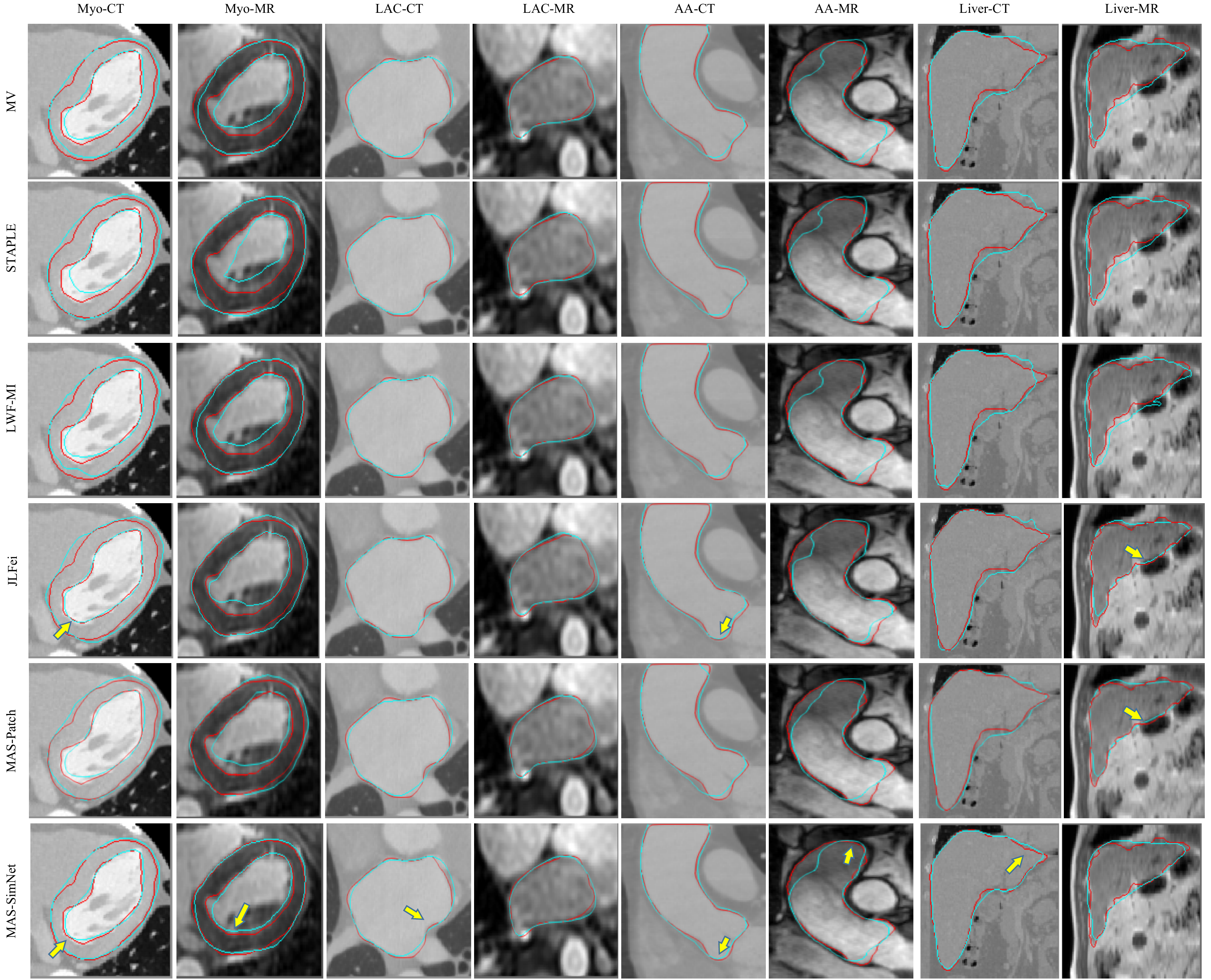}
		\caption{ Visualization of Myo and liver segmentation results. 
			The showed images are the median cases in terms of Dice score by MAS-SimNet.  
			\hlo{The gold standard and predicted labels are delineated by red and cyan contours, respectively. Arrows point to the areas where MAS-Patch, JLF\hlo{$_{ei}$} and MAS-SimNet achieved better performances than other label fusion methods.
			}  
		}
		\label{fig:seg_mas}    
	\end{figure*}

	Additionally, Fig. \ref{fig:seg_mas} shows four representative segmentation cases from the two datasets. 
	Although all the label fusion methods generally obtained good results inside the anatomies, while MAS-SimNet, \hlo{MAS-Patch} and JLF\hlo{$_{ei}$} provided relatively better details on the boundaries.
	
\begin{figure}[htp]
		\centering
		\includegraphics[width=0.85\columnwidth]{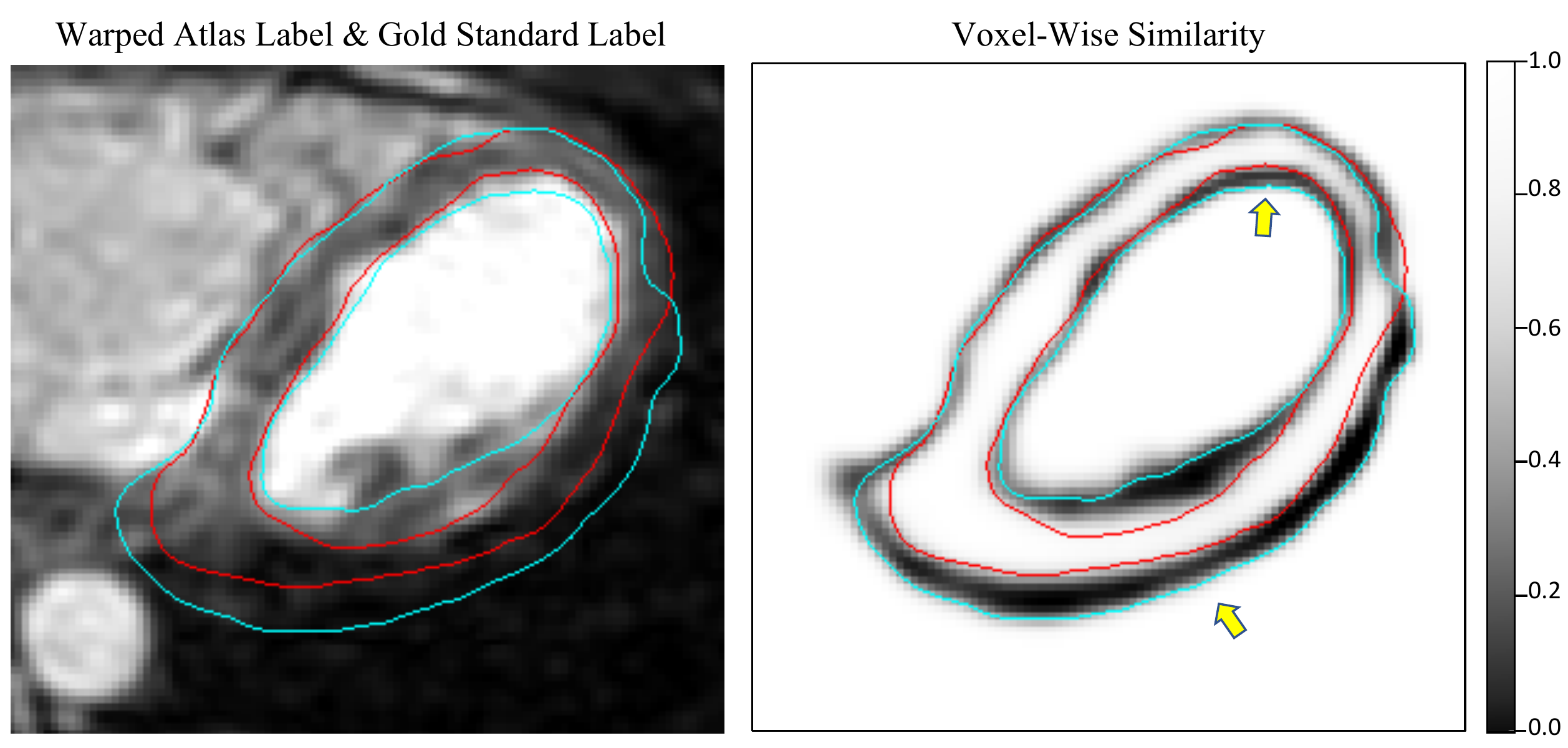}
		\caption{Visualization of an estimated similarity. 			
			In the voxel-wise similarity map, the lower similarity values  ($\approx$ 0) are in dark, and the higher similarity values  ($\approx$ 1.0) are in bright. 
			The arrows point to several inconsistent areas of the warped atlas label (cyan contour) when comparing to the target gold standard label (red contour). 
		}
		\label{fig:sim_visaul}    
	\end{figure}

	Furthermore, Fig. \ref{fig:sim_visaul} visualizes a representative voxel-wise similarity from SimNet.
	Here, the warped atlas label is one of the candidate segmentations of the target image. 
	One can see that SimNet could well assign low similarity values (dark color) for the incorrect regions, and preserved high confidences (bright color) for the correct areas. 
	Hence, the estimated similarity by SimNet can be adopted for the label fusion step effectively.

	\subsubsection{Effectiveness of Cross-Modality Atlases}
	\label{sec:res:seg}
	The modality and quality of the atlases could potentially impact the accuracy of MAS method. We compared MR$\rightarrow$MR (using MR atlases to segment MR targets) to  CT$\rightarrow$MR (using CT atlases to segment MR targets, see Table \ref{tab:reg_result} and \ref{tab:mas}) to investigate the benefit of cross-modality atlases. Table \ref{tab:mr} summarizes the registration (BiRegNet) and label fusion (MAS-SimNet) results under MR$\rightarrow$MR setup. After the label fusion step, the DS of CT$\rightarrow$MR were close to MR$\rightarrow$MR on LVC (90.78\% VS. 91.16\%, $p=0.216$) and Myo (81.06\% VS. 81.03\%, $p=0.953$) structures. This indicates that the cross-modality (CT) atlases could obtain comparable result to the mono-modality (MR) atlases. However, for liver segmentation, CT$\rightarrow$MR achieved statistically better (91.24\% VS. 90.70\%, $p<0.05$) DS result against MR$\rightarrow$MR. This is reasonable, as the original coronal resolution of MR ($5.5\sim9$ mm) is much lower than that of CT ($3\sim3.2$ mm) in CHAOS dataset (see Fig \ref{fig:0} (a)). Thus, CT$\rightarrow$MR which employed high quality (resolution) CT atlases could be better than MR$\rightarrow$MR. 
	\begin{table}[htp]
		\centering
		\caption{The results of registration and label fusion under MR$\rightarrow$MR setup. We performed the registration and label fusion steps via BiRegNet and MAS-SimNet, respectively.}
		\resizebox{0.5\textwidth}{!}{
			\begin{tabular}{l|ll|ll}
				\hline
				\multirow{2}[0]{*}{Target} & \multicolumn{2}{c|}{BiRegNet} & \multicolumn{2}{c}{MAS-SimNet}  \\
				\cline{2-5}
				& \multicolumn{1}{c}{DS (\%)$\uparrow$ } & \multicolumn{1}{c|}{ASD (mm)$\downarrow$} & \multicolumn{1}{c}{DS (\%)$\uparrow$} & \multicolumn{1}{c}{ASD (mm)$\downarrow$} \\
				\hline
				LVC      & {88.82$\pm$4.30}  & {1.81$\pm$0.50}  & {91.16$\pm$3.50}     & {1.45$\pm$0.41}  \\
				Myo       & {73.30$\pm$5.59}  & {2.03$\pm$0.52}  & {81.03$\pm$3.89}    & {1.55$\pm$0.41} \\
				\hdashline
				Liver    & {88.15$\pm$1.66} & {3.90$\pm$0.64} & {90.70$\pm$1.25}     & {3.06$\pm$0.34} \\
				\hline
			\end{tabular}%
		}
		\label{tab:mr}%
	\end{table}%

	\subsubsection{Study of the Number of Atlases}
\begin{figure}[htp]
	    \centering
	    \includegraphics[width=\columnwidth]{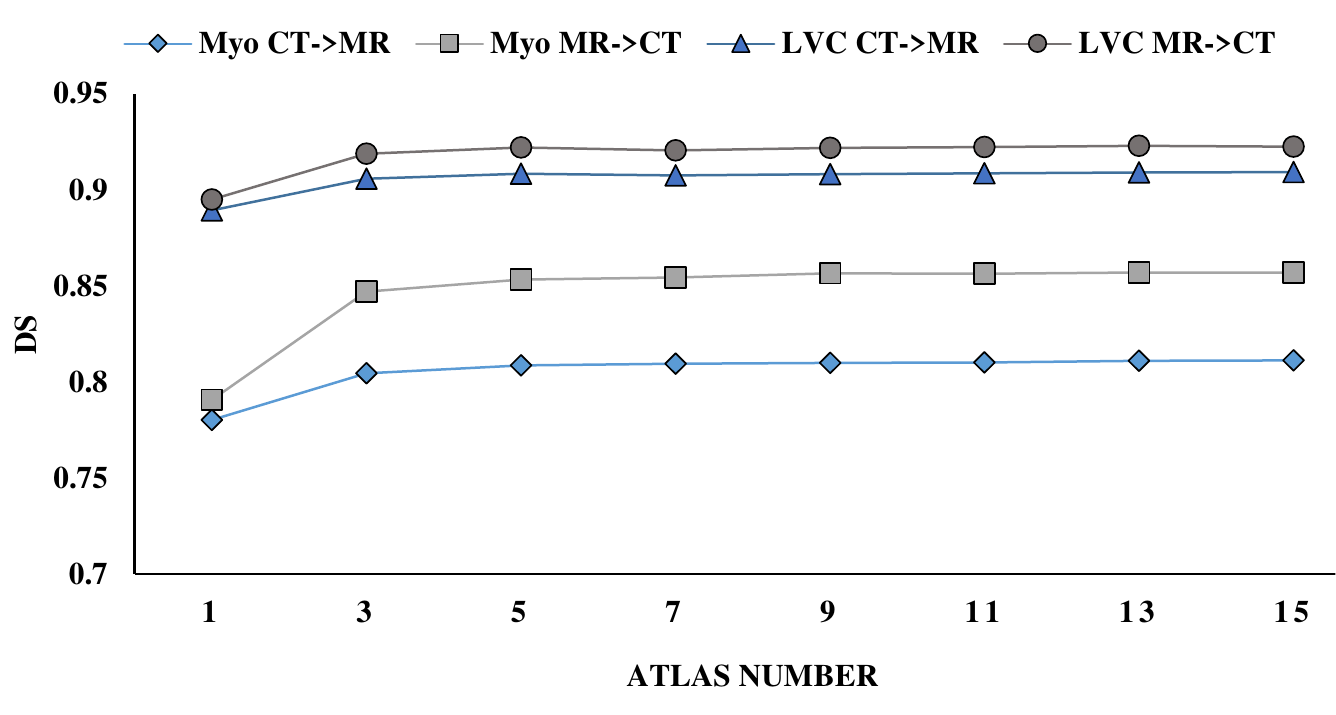}
	    \caption{\hlo{The results of using different numbers of atlases for Myo  and LVC segmentation. Both CT$\rightarrow$MR and MR$\rightarrow$CT setups are included. 
	    }}
	    \label{fig:main:atlas_num}
	\end{figure}

	\hlo{The number of atlases could affect the performance of MAS methods \cite{jour/tmi/bai2013probabilistic}. In this study, we investigated the effect of the number of atlases on the performance of MAS-SimNet for LV segmentation. As shown in Figure \ref{fig:main:atlas_num}, the performance of MAS-SimNet initially improved  along with the increase of the number of atlases, and then gradually converged after using seven atlases. This is consistent to the trend reported in current mono-modality MAS methods \cite{jour/tmi/bai2013probabilistic,jour/ni/aljabar2009multi}.}
	
	\section{Discussion and Conclusion}
	\label{sec:diss}
	

	This paper presents an automatic deep learning based cross-modality MAS framework.      
	The motivation is to illustrate the potential and performance of utilizing cross-modality atlases via DNNs in MAS framework. Two major contributions have been introduced. 
	The first is BiRegNet which can jointly estimate a pair of forward and backward DDFs for a registration. 
	In BiRegNet, we further propose a consistency loss for realistic DDFs, as an alternative regularization term for registration networks \cite{conf/miccai/luo2020mvmm,journal/mia/hu2018weakly}. 
	Besides, the attention-based module, \ie, AG, could also improve the performance of BiRegNet. 
	The second is the cross-modality multi-atlas fusion method, where the fusion weights (similarities) are estimated by a DNN model, \ie, SimNet. 
	To obtain a better similarity estimation, SimNet is designed to aggregate multi-scale information for the similarity estimation. 
	The framework has been applied to two clinical tasks, i.e., LV segmentation and liver segmentation. 
	Results have shown the effectiveness of the proposed DNN-based registration (see Table \ref{tab:reg_result}) and label fusion (see Table \ref{tab:mas}) methods. 
	For the computation time, it costed average 0.9 seconds and 8.3 seconds for the registration and label fusion in our experiments, respectively.

	In the literature, several works have been proposed to use cross-modality atlases for segmentation. Iglesias \etal \cite{journal/MIA/EugenioIglesias2013} presented a generative probabilistic model for solving the cross-modality registration and label fusion steps. 
	They used 20 proton density MR atlases to segment 8 T1-weighted MR brain images and vice versa. The model significantly outperformed MV method. Kasiri \etal \cite{conf/icip/kasiri2014cross} designed a similarity based on un-decimated wavelet transform for cross-modality atlas fusion. They achieved T2 MR brain targets segmentation using T1 MR atlases, and obtained average DS for white matter, grey matter and cerebrospinal fluid of the brain were 80.6\%, 75.0\% and 61.2\%, respectively. Furthermore, Zhuang \& Shen \cite{journal/mia/zhuang2016mmm} proposed a multi-modality MAS framework, and achieved	85.0\% and 94.0\% DS for Myo and LVC segmentation on MR images. 
	Such results are better than the proposed method (see Table \ref{tab:mas}). 
	The main reason is that they performed a cardiac MR image segmentation by jointly using same-modality (MR) and cross-modality (CT) atlases, whereas our work focus on utilizing the cross-modality (CT) ones.      
	
	At present, semantic segmentation methods are the most widely studied methods for medical image segmentation. These methods resolve segmentation problem via pixel-level classification. Table \ref{tab:semantic} summaries the top performances of semantic segmentation networks on MM-WHS and CHAOS datasets. Semantic segmentation methods (Seg-CNN, PKDIA or nnUNet ) won the best places on the corresponding challenge of the datasets. \hlo{Particularly, nnUNet and PKDIA demonstrated effectiveness on CHAOS dataset. It may be due to their employed data augmentation techniques (Gaussian noise/blur, simulated low resolution and Gamma transform) they employed.} However, for MM-WHS, compared to the winner of MM-WHS (Seg-CNN), the atlas-based segmentation method MAS-SimNet could achieve almost 6\% (81.1\% VS. 75.2\%) DS improvement for Myo segmentation on MR images, \hlo{thanks to the advantage of using the high quality CT images as atlases.} Note that Myo segmentation is more difficult than LVC and liver segmentation, and MR is considered as more challenging (ambiguity boundaries, motion artifacts and noises) data compared to CT. This proved the robustness of MAS-SimNet. 

	\begin{table}[htp] 
	\centering
		\caption{\coloro{DS (\%) of different state-of-the art semantic segmentation methods on the MM-WHS and CHAOS dataset. The quantitative results of Seg-CNN, 3D-FCN, eFCN, nnUNet, PKDIA, Mountain, MedianCHAOS5 and ISDUE are directly adopted from \cite{conf/miccai-ws/payer2017multi,conf/miccai-ws/yang20173d,conf/miccai-ws/yang2017hybrid,joural/mia/kavur2020chaos}. Please also note that these semantic segmentation methods were conducted to train and evaluate on a mono-modality (CT or MR) of medical images. $^\star$ denotes the winners of the challenges.}}
		\centering
		
		\resizebox{\columnwidth}{!}{
						\begin{tabular}{c|l|ll||ll}   
				\hline   
				&\hlo{Method}  & \hlo{LVC}  & \hlo{Myo} & \hlo{Method} & \hlo{Liver} \\ 
				\hline
				\multirow{4}{*}{\rotatebox{90}{\hlo{MR}}} 
				
				& \hlo{Seg-CNN$^\star$ \cite{conf/miccai-ws/payer2017multi}} & \hlo{87.7} &\hlo{75.2}   & \hlo{nnUNet$^\star$ \cite{joural/mia/kavur2020chaos}}  	&	  \textbf{\hlo{95.0}}\\  
				
				& \hlo{3D-FCN \cite{conf/miccai-ws/yang20173d}} & \hlo{87.1}& \hlo{71.9}&  \hlo{PKDIA\cite{joural/mia/kavur2020chaos}}  &    \hlo{\underline{94.0}} \\      
				
				&  \hlo{eFCN \cite{conf/miccai-ws/yang2017hybrid}} & \hlo{85.8} & \hlo{74.2} &  \hlo{Mountain\cite{joural/mia/kavur2020chaos}}   &   \hlo{92.0} \\ 
	
							\hdashline
				\hlo{CT$\rightarrow$ MR} & \hlo{MAS-SimNet}   & \hlo{\textbf{90.8}}	&  \hlo{\textbf{81.1}} & \hlo{MAS-SimNet}   & \hlo{91.2}   \\  
				\hline
				\hline
				
				\multirow{4}{*}{\rotatebox{90}{\hlo{CT}}} 
				  
				& \hlo{Seg-CNN$^\star$ \cite{conf/miccai-ws/payer2017multi}} & \hlo{\textbf{92.4}} & \hlo{\textbf{87.2}} & \hlo{PKDIA$^\star$ \cite{joural/mia/kavur2020chaos}}	&	 \hlo{\textbf{98.0}} \\
			
				
				&  \hlo{3D-FCN \cite{conf/miccai-ws/yang20173d}} & \hlo{90.8}& \hlo{81.3}	& \hlo{MedianCHAOS5 \cite{joural/mia/kavur2020chaos}}  		&	 \hlo{\underline{97.0}}\\ 
				
				&  \hlo{eFCN \cite{conf/miccai-ws/yang2017hybrid}} & \hlo{87.8}& \hlo{81.8} & \hlo{ISDUE\cite{joural/mia/kavur2020chaos}}  	&	 \hlo{91.0} \\  
				

							\hdashline	
				\hlo{MR$\rightarrow$ CT}	& \hlo{MAS-SimNet}   & \hlo{\underline{91.9}}	&  \hlo{\underline{85.7}} & \hlo{MAS-SimNet}   &  \hlo{92.2}  \\  
				\hline
			\end{tabular} 
		}
		\label{tab:semantic}
	\end{table}

	There are four limitations of this work. 
	First, both BiRegNet and SimNet are trained with anatomical labels in our framework, which limits the framework to the unlabelled datasets. 
	Hence, one direction of future work is extending the framework with unsupervised registration and label fusion DNN models, \hlo{which could also be a potential solution for cross-domain image segmentation \cite{jour/tmi/chen2020unsupervised,conf/icml/hoffman2018cycada}.}
	Second, BiRegNet and SimNet are optimized separately, which increases the training complication of the framework. 
	Thus, it is worth to combine the registration and fusion DNN models together to achieve an end-to-end framework in future. 
	\hlo{Third}, the method only uses cross-modality atlases for segmentation, while multi-modality images are available in some clinical segmentation tasks.
	For example, Zhuang \cite{axiv/zhuang2020cardiac} provided three modalities of cardiac MR images, \ie, LGE, bSSFP and T2, which makes it possible to construct multi-modality atlases for segmentation \hlo{in the future. 
Meanwhile, several opening questions, such as how to appropriately utilize multi-modality atlases to substantially improve the performance of segmentation, still remain to be explored and verified.}
\hlo{Finally, our method processes 3D medical images which have high demand of computational resources, and thus we only crop a small field-of-view (FOV) as input.
However, sometimes we need to segment multiple organs simultaneously from certain medical images, such as abdominal images.
It is valuable to explore new memory-efficient network architectures \cite{conf/miccai/heinrich2020highly} for MAS methods in the future.}
	
\bibliographystyle{ieeetr}
\bibliography{refs}

\begin{thebibliography}{10}

\bibitem{jour/jbhi/xue2021left}
W.~Xue, J.~Li, Z.~Hu, E.~Kerfoot, J.~Clough, I.~Oksuz, H.~Xu, V.~Grau, F.~Guo,
  M.~Ng, {\em et~al.}, ``Left ventricle quantification challenge: A
  comprehensive comparison and evaluation of segmentation and regression for
  mid-ventricular short-axis cardiac mr data,'' {\em IEEE Journal of Biomedical
  and Health Informatics}, 2021.

\bibitem{jour/jbhi/ji2013acm}
H.~Ji, J.~He, X.~Yang, R.~Deklerck, and J.~Cornelis, ``{ACM}-based automatic
  liver segmentation from 3-{D} {CT} images by combining multiple atlases and
  improved mean-shift techniques,'' {\em IEEE Journal of Biomedical and Health
  Informatics}, vol.~17, no.~3, pp.~690--698, 2013.

\bibitem{journal/mia/iglesias2015multi}
J.~E. Iglesias and M.~R. Sabuncu, ``Multi-atlas segmentation of biomedical
  images: a survey,'' {\em Medical image analysis}, vol.~24, no.~1,
  pp.~205--219, 2015.

\bibitem{journal/mia/antonelli2019gas}
M.~Antonelli, M.~J. Cardoso, E.~W. Johnston, M.~B. Appayya, B.~Presles,
  M.~Modat, S.~Punwani, and S.~Ourselin, ``Gas: A genetic atlas selection
  strategy in multi-atlas segmentation framework,'' {\em Medical image
  analysis}, vol.~52, pp.~97--108, 2019.

\bibitem{journal/PAMI/kittler1998combining}
J.~Kittler, M.~Hatef, R.~P. Duin, and J.~Matas, ``On combining classifiers,''
  {\em IEEE transactions on pattern analysis and machine intelligence},
  vol.~20, no.~3, pp.~226--239, 1998.

\bibitem{journal/tmi/Tang2019}
Z.~Tang, P.-T. Yap, and D.~Shen, ``A new multi-atlas registration framework for
  multimodal pathological images using conventional monomodal normal atlases,''
  {\em IEEE Transactions on Image Processing}, vol.~28, no.~5, pp.~2293--2304,
  2018.

\bibitem{joural/mia/kavur2020chaos}
A.~E. Kavur, N.~S. Gezer, M.~Bar{\i}{\c{s}}, S.~Aslan, P.-H. Conze, V.~Groza,
  D.~D. Pham, S.~Chatterjee, P.~Ernst, S.~{\"O}zkan, {\em et~al.}, ``{CHAOS}
  challenge-combined ({CT-MR}) healthy abdominal organ segmentation,'' {\em
  Medical Image Analysis}, p.~101950, 2020.

\bibitem{jour/mia/li2020atrial}
L.~Li, F.~Wu, G.~Yang, L.~Xu, T.~Wong, R.~Mohiaddin, D.~Firmin, J.~Keegan, and
  X.~Zhuang, ``Atrial scar quantification via multi-scale cnn in the graph-cuts
  framework,'' {\em Medical image analysis}, vol.~60, p.~101595, 2020.

\bibitem{journal/MIA/EugenioIglesias2013}
J.~E. Iglesias, M.~R. Sabuncu, and K.~Van~Leemput, ``A unified framework for
  cross-modality multi-atlas segmentation of brain {MRI},'' {\em Medical image
  analysis}, vol.~17, no.~8, pp.~1181--1191, 2013.

\bibitem{conf/icip/kasiri2014cross}
K.~Kasiri, P.~Fieguth, and D.~A. Clausi, ``Cross modality label fusion in
  multi-atlas segmentation,'' in {\em 2014 IEEE International Conference on
  Image Processing (ICIP)}, pp.~16--20, IEEE, 2014.

\bibitem{jour/tmi/bai2013probabilistic}
W.~Bai, W.~Shi, D.~P. O'regan, T.~Tong, H.~Wang, S.~Jamil-Copley, N.~S. Peters,
  and D.~Rueckert, ``A probabilistic patch-based label fusion model for
  multi-atlas segmentation with registration refinement: application to cardiac
  mr images,'' {\em IEEE transactions on medical imaging}, vol.~32, no.~7,
  pp.~1302--1315, 2013.

\bibitem{journal/mia/sanroma2018lf}
G.~Sanroma, O.~M. Benkarim, G.~Piella, O.~Camara, G.~Wu, D.~Shen, J.~D.
  Gispert, J.~L. Molinuevo, M.~A.~G. Ballester, A.~D.~N. Initiative, {\em
  et~al.}, ``Learning non-linear patch embeddings with neural networks for
  label fusion,'' {\em Medical image analysis}, vol.~44, pp.~143--155, 2018.

\bibitem{conf/miccai/coup2010}
P.~Coup{\'e}, J.~V. Manj{\'o}n, V.~Fonov, J.~Pruessner, M.~Robles, and D.~L.
  Collins, ``Nonlocal patch-based label fusion for hippocampus segmentation,''
  in {\em International Conference on Medical Image Computing and
  Computer-Assisted Intervention}, pp.~129--136, Springer, 2010.

\bibitem{arxiv/Oktay}
O.~Oktay, J.~Schlemper, L.~L. Folgoc, M.~Lee, M.~Heinrich, K.~Misawa, K.~Mori,
  S.~McDonagh, N.~Y. Hammerla, B.~Kainz, {\em et~al.}, ``Attention u-net:
  {L}earning where to look for the pancreas,'' in {\em International Conference
  on Medical Imaging with Deep Learning}, 2018.
\newblock [Online] \url{https://openreview.net/forum?id=Skft7cijM}.

\bibitem{journal/mp/Zhuang2015}
X.~Zhuang, W.~Bai, J.~Song, S.~Zhan, X.~Qian, W.~Shi, Y.~Lian, and D.~Rueckert,
  ``Multiatlas whole heart segmentation of {CT} data using conditional entropy
  for atlas ranking and selection,'' {\em Medical physics}, vol.~42, no.~7,
  pp.~3822--3833, 2015.

\bibitem{journal/mia/zhuang2019evaluation}
X.~Zhuang, L.~Li, C.~Payer, D.~{\v{S}}tern, M.~Urschler, M.~P. Heinrich,
  J.~Oster, C.~Wang, {\"O}.~Smedby, C.~Bian, {\em et~al.}, ``Evaluation of
  algorithms for multi-modality whole heart segmentation: an open-access grand
  challenge,'' {\em Medical image analysis}, vol.~58, p.~101537, 2019.

\bibitem{conf/miccai/ding2020cross}
W.~Ding, L.~Li, X.~Zhuang, and L.~Huang, ``Cross-modality multi-atlas
  segmentation using deep neural networks,'' in {\em International Conference
  on Medical Image Computing and Computer-Assisted Intervention}, pp.~233--242,
  Springer, 2020.

\bibitem{journal/mia/zhuang2016mmm}
X.~Zhuang and J.~Shen, ``Multi-scale patch and multi-modality atlases for whole
  heart segmentation of {MRI},'' {\em Medical image analysis}, vol.~31,
  pp.~77--87, 2016.

\bibitem{journal/tmi/warfield2004simultaneous}
S.~K. Warfield, K.~H. Zou, and W.~M. Wells, ``Simultaneous truth and
  performance level estimation ({STAPLE}): an algorithm for the validation of
  image segmentation,'' {\em IEEE transactions on medical imaging}, vol.~23,
  no.~7, pp.~903--921, 2004.

\bibitem{jour/pami/wang2012multi}
H.~Wang, J.~W. Suh, S.~R. Das, J.~B. Pluta, C.~Craige, and P.~A. Yushkevich,
  ``Multi-atlas segmentation with joint label fusion,'' {\em IEEE transactions
  on pattern analysis and machine intelligence}, vol.~35, no.~3, pp.~611--623,
  2012.

\bibitem{conf/miccai/ding2019votenet}
Z.~Ding, X.~Han, and M.~Niethammer, ``Vote{N}et: A deep learning label fusion
  method for multi-atlas segmentation,'' in {\em International Conference on
  Medical Image Computing and Computer-Assisted Intervention}, pp.~202--210,
  Springer, 2019.

\bibitem{conf/miccai/xie2019msg}
L.~Xie, J.~Wang, M.~Dong, D.~A. Wolk, and P.~A. Yushkevich, ``Improving
  multi-atlas segmentation by convolutional neural network based patch error
  estimation,'' in {\em International Conference on Medical Image Computing and
  Computer-Assisted Intervention}, pp.~347--355, Springer, 2019.

\bibitem{journal/mia/yang2018neural}
H.~Yang, J.~Sun, H.~Li, L.~Wang, and Z.~Xu, ``Neural multi-atlas label fusion
  application to cardiac {MR} images,'' {\em Medical image analysis}, vol.~49,
  pp.~60--75, 2018.

\bibitem{journal/pr/luan2008multimodality}
H.~Luan, F.~Qi, Z.~Xue, L.~Chen, and D.~Shen, ``Multimodality image
  registration by maximization of quantitative--qualitative measure of mutual
  information,'' {\em Pattern Recognition}, vol.~41, no.~1, pp.~285--298, 2008.

\bibitem{journal/pr/studholme1999overlap}
C.~Studholme, D.~L. Hill, and D.~J. Hawkes, ``An overlap invariant entropy
  measure of 3{D} medical image alignment,'' {\em Pattern recognition},
  vol.~32, no.~1, pp.~71--86, 1999.

\bibitem{journal/tmi/zhuang2011nonrigid}
X.~Zhuang, S.~Arridge, D.~J. Hawkes, and S.~Ourselin, ``A nonrigid registration
  framework using spatially encoded mutual information and free-form
  deformations,'' {\em IEEE transactions on medical imaging}, vol.~30, no.~10,
  pp.~1819--1828, 2011.

\bibitem{conf/miccai/qin2019unsupervised}
C.~Qin, B.~Shi, R.~Liao, T.~Mansi, D.~Rueckert, and A.~Kamen, ``Unsupervised
  deformable registration for multi-modal images via disentangled
  representations,'' in {\em International Conference on Information Processing
  in Medical Imaging}, pp.~249--261, Springer, 2019.

\bibitem{journal/mia/wachinger2012entropy}
C.~Wachinger and N.~Navab, ``Entropy and {L}aplacian images: {S}tructural
  representations for multi-modal registration,'' {\em Medical image analysis},
  vol.~16, no.~1, pp.~1--17, 2012.

\bibitem{journal/mia/Heinrich2012}
M.~P. Heinrich, M.~Jenkinson, M.~Bhushan, T.~Matin, F.~V. Gleeson, M.~Brady,
  and J.~A. Schnabel, ``{MIND}: {M}odality independent neighbourhood descriptor
  for multi-modal deformable registration,'' {\em Medical image analysis},
  vol.~16, no.~7, pp.~1423--1435, 2012.

\bibitem{journal/mva/haskins2020deep}
G.~Haskins, U.~Kruger, and P.~Yan, ``Deep learning in medical image
  registration: a survey,'' {\em Machine Vision and Applications}, vol.~31,
  no.~1, pp.~1--18, 2020.

\bibitem{journal/mia/Fan2019}
J.~Fan, X.~Cao, P.-T. Yap, and D.~Shen, ``{BIRN}et: {B}rain image registration
  using dual-supervised fully convolutional networks,'' {\em Medical image
  analysis}, vol.~54, pp.~193--206, 2019.

\bibitem{journal/ni/Yang2017}
X.~Yang, R.~Kwitt, M.~Styner, and M.~Niethammer, ``Quicksilver: {F}ast
  predictive image registration--a deep learning approach,'' {\em NeuroImage},
  vol.~158, pp.~378--396, 2017.

\bibitem{journal/tmi/Balakrishnan2019}
G.~Balakrishnan, A.~Zhao, M.~R. Sabuncu, J.~Guttag, and A.~V. Dalca,
  ``Voxelmorph: a learning framework for deformable medical image
  registration,'' {\em IEEE transactions on medical imaging}, vol.~38, no.~8,
  pp.~1788--1800, 2019.

\bibitem{journal/mia/hu2018weakly}
Y.~Hu, M.~Modat, E.~Gibson, W.~Li, N.~Ghavami, E.~Bonmati, G.~Wang, S.~Bandula,
  C.~M. Moore, M.~Emberton, {\em et~al.}, ``Weakly-supervised convolutional
  neural networks for multimodal image registration,'' {\em Medical image
  analysis}, vol.~49, pp.~1--13, 2018.

\bibitem{journal/tmi/christensen2001consis}
G.~E. Christensen and H.~J. Johnson, ``Consistent image registration,'' {\em
  IEEE transactions on medical imaging}, vol.~20, no.~7, pp.~568--582, 2001.

\bibitem{conf/miccai/kim2019unsupervised}
B.~Kim, J.~Kim, J.-G. Lee, D.~H. Kim, S.~H. Park, and J.~C. Ye, ``Unsupervised
  deformable image registration using cycle-consistent cnn,'' in {\em
  International Conference on Medical Image Computing and Computer-Assisted
  Intervention}, pp.~166--174, Springer, 2019.

\bibitem{conf/cvpr/mok2020fast}
T.~C. Mok and A.~Chung, ``Fast symmetric diffeomorphic image registration with
  convolutional neural networks,'' in {\em Proceedings of the IEEE/CVF
  conference on computer vision and pattern recognition}, pp.~4644--4653, 2020.

\bibitem{conf/miccai/gu2020pair}
D.~Gu, X.~Cao, S.~Ma, L.~Chen, G.~Liu, D.~Shen, and Z.~Xue, ``Pair-wise and
  group-wise deformation consistency in deep registration network,'' in {\em
  International Conference on Medical Image Computing and Computer-Assisted
  Intervention}, pp.~171--180, Springer, 2020.

\bibitem{journal/tmi/artaechevarria2009combination}
X.~Artaechevarria, A.~Munoz-Barrutia, and C.~Ortiz-de Sol{\'o}rzano,
  ``Combination strategies in multi-atlas image segmentation: application to
  brain {MR} data,'' {\em IEEE transactions on medical imaging}, vol.~28,
  no.~8, pp.~1266--1277, 2009.

\bibitem{jouranl/mia/dou20173d}
Q.~Dou, L.~Yu, H.~Chen, Y.~Jin, X.~Yang, J.~Qin, and P.-A. Heng, ``3d deeply
  supervised network for automated segmentation of volumetric medical images,''
  {\em Medical image analysis}, vol.~41, pp.~40--54, 2017.

\bibitem{journal/ni/klein2009evaluation}
A.~Klein, J.~Andersson, B.~A. Ardekani, J.~Ashburner, B.~Avants, M.-C. Chiang,
  G.~E. Christensen, D.~L. Collins, J.~Gee, P.~Hellier, {\em et~al.},
  ``Evaluation of 14 nonlinear deformation algorithms applied to human brain
  {MRI} registration,'' {\em Neuroimage}, vol.~46, no.~3, pp.~786--802, 2009.

\bibitem{journal/inj/avants2009advanced}
B.~B. Avants, N.~Tustison, and G.~Song, ``Advanced normalization tools
  ({ANTS}),'' {\em Insight j}, vol.~2, no.~365, pp.~1--35, 2009.

\bibitem{jour/ni/aljabar2009multi}
P.~Aljabar, R.~A. Heckemann, A.~Hammers, J.~V. Hajnal, and D.~Rueckert,
  ``Multi-atlas based segmentation of brain images: atlas selection and its
  effect on accuracy,'' {\em Neuroimage}, vol.~46, no.~3, pp.~726--738, 2009.

\bibitem{conf/miccai/luo2020mvmm}
X.~Luo and X.~Zhuang, ``Mv{MM}-{R}egnet: {A} new image registration framework
  based on multivariate mixture model and neural network estimation,'' in {\em
  International Conference on Medical Image Computing and Computer-Assisted
  Intervention}, pp.~149--159, Springer, 2020.

\bibitem{conf/miccai-ws/payer2017multi}
C.~Payer, D.~{\v{S}}tern, H.~Bischof, and M.~Urschler, ``Multi-label whole
  heart segmentation using cnns and anatomical label configurations,'' in {\em
  International Workshop on Statistical Atlases and Computational Models of the
  Heart}, pp.~190--198, Springer, 2017.

\bibitem{conf/miccai-ws/yang20173d}
X.~Yang, C.~Bian, L.~Yu, D.~Ni, and P.-A. Heng, ``3{D} convolutional networks
  for fully automatic fine-grained whole heart partition,'' in {\em
  International Workshop on Statistical Atlases and Computational Models of the
  Heart}, pp.~181--189, Springer, 2017.

\bibitem{conf/miccai-ws/yang2017hybrid}
X.~Yang, C.~Bian, L.~Yu, D.~Ni, and P.-A. Heng, ``Hybrid loss guided
  convolutional networks for whole heart parsing,'' in {\em International
  workshop on statistical atlases and computational models of the heart},
  pp.~215--223, Springer, 2017.

\bibitem{jour/tmi/chen2020unsupervised}
C.~Chen, Q.~Dou, H.~Chen, J.~Qin, and P.~A. Heng, ``Unsupervised bidirectional
  cross-modality adaptation via deeply synergistic image and feature alignment
  for medical image segmentation,'' {\em IEEE transactions on medical imaging},
  vol.~39, pp.~2494--2505, Feb 2020.

\bibitem{conf/icml/hoffman2018cycada}
J.~Hoffman, E.~Tzeng, T.~Park, J.-Y. Zhu, P.~Isola, K.~Saenko, A.~Efros, and
  T.~Darrell, ``Cycada: Cycle-consistent adversarial domain adaptation,'' in
  {\em International conference on machine learning}, pp.~1989--1998, PMLR,
  2018.

\bibitem{axiv/zhuang2020cardiac}
X.~Zhuang, J.~Xu, X.~Luo, C.~Chen, C.~Ouyang, D.~Rueckert, V.~M. Campello,
  K.~Lekadir, S.~Vesal, N.~RaviKumar, {\em et~al.}, ``Cardiac segmentation on
  late gadolinium enhancement {MRI}: a benchmark study from multi-sequence
  cardiac {MR} segmentation challenge,'' {\em arXiv preprint arXiv:2006.12434},
  2020.

\bibitem{conf/miccai/heinrich2020highly}
M.~P. Heinrich and L.~Hansen, ``Highly accurate and memory efficient
  unsupervised learning-based discrete ct registration using 2.5 {D}
  displacement search,'' in {\em International Conference on Medical Image
  Computing and Computer-Assisted Intervention}, pp.~190--200, Springer, 2020.

\end{thebibliography}
	
\end{document}


\title{Supplementary Material}
	\author{Wangbin Ding, Lei Li, Xiahai Zhuang*, and Liqin Huang*
		\thanks{
		}
		\thanks{}
		\thanks{}
		\thanks{}
	}
	
	\maketitle

\begin{figure*}[htp]
    \centering
    \vspace{-1cm}
    \setlength{\abovecaptionskip}{3pt}
    \setlength{\belowcaptionskip}{5pt}
    \includegraphics[width=0.84\textwidth]{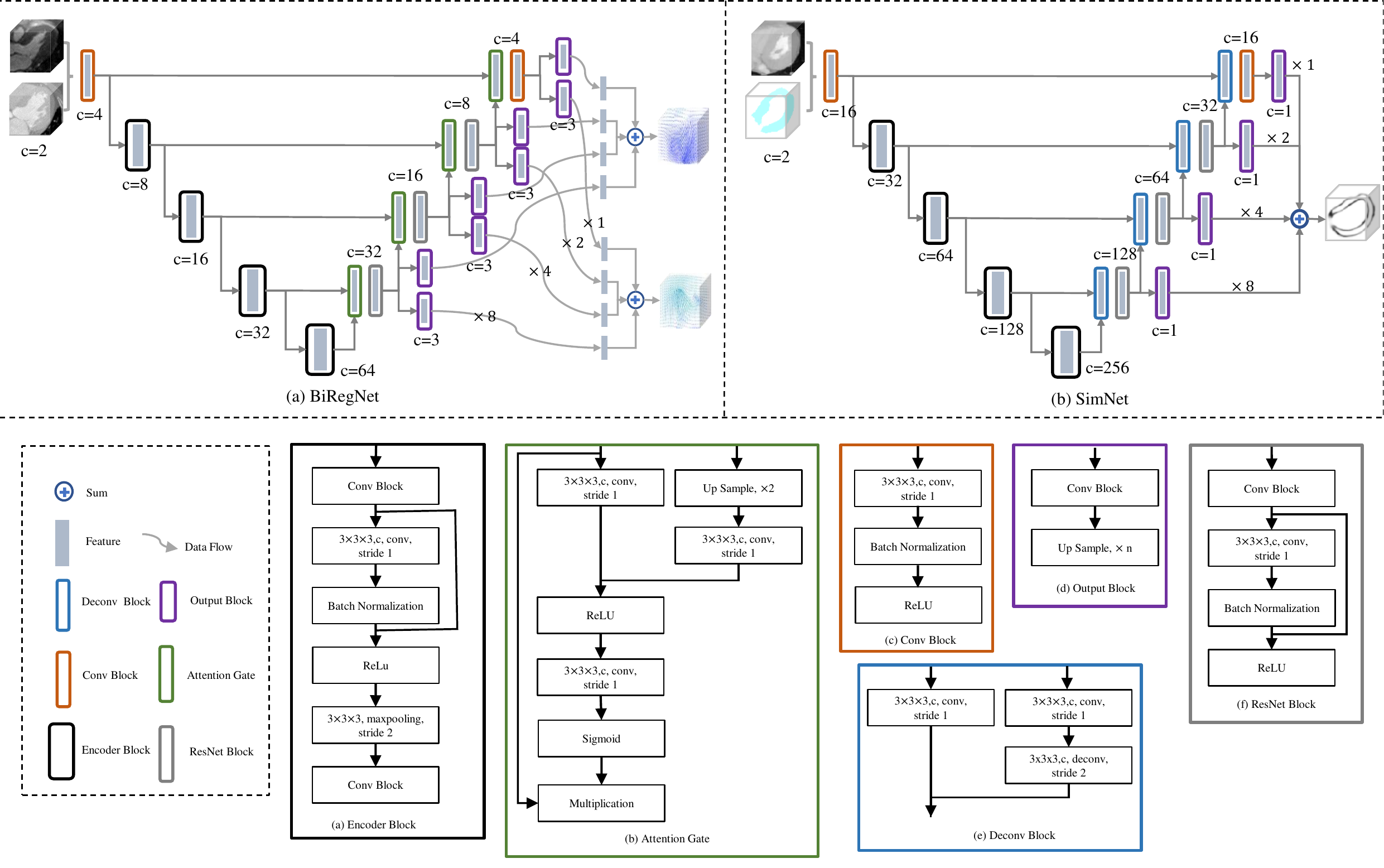}
    \caption{\hlo{(a) The network architecture of BiRegNet. It mainly includes four encoders (black) and four decoders (green \& gray). Each decoder contains an attention gate (green) and a ResNet block (gray). At output layers, we up-sample different resolution levels of feature maps into an unified size, and concatenate them together to predict forward and backward DDFs. (b) The network architecture of SimNet, which contains four encoders (black) and four decoders (blue \& gray). We up-sample and aggregate multiple levels of feature maps when performing voxel-level similarity estimation.}}
    \label{fig:sup:arch}
\end{figure*}

\begin{figure*}[htp]
    \centering
    \vspace{-0.6cm}
    \setlength{\abovecaptionskip}{3pt}
    \setlength{\belowcaptionskip}{0pt}
    \includegraphics[width=0.85\textwidth]{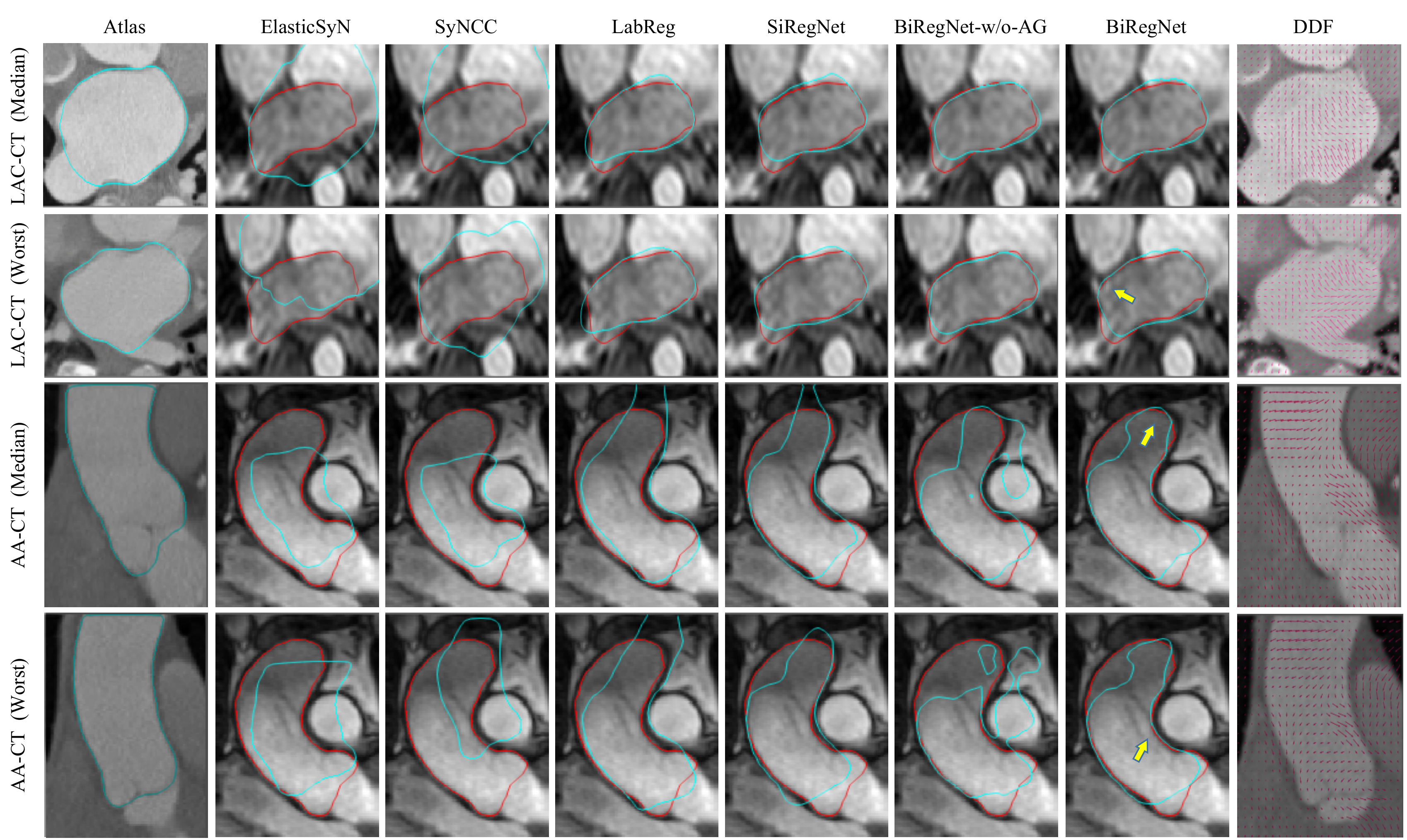}
    \caption{\hlo{Visualization of LAC and AA registration results. The presented images are the median and worst cases in terms of Dice score of BiRegNet. The red contours delineate the gold standard labels, and the cyan contours delineate the (warped) atlas labels. The arrows point to the regions where BiRegNet achieves better visual results than the other registration methods. The dense arrows in last column visualize the DDF of BiRegNet. (The reader is referred to the colorful web version of this article.) .}}
    \label{fig:reg_ressult}
\end{figure*}